\documentclass[acmsmall]{acmart}

\usepackage{hyperref}

\usepackage{hyperxmp}

\usepackage{amsmath,amsfonts}
\usepackage{algorithmic}
\usepackage{graphicx}
\usepackage{textcomp}
\usepackage{xcolor}
\usepackage{booktabs}
\definecolor{Gray}{gray}{0.9}
\usepackage{tabularx}
\usepackage{makecell}
\usepackage{verbatim}

\usepackage{multirow}
\usepackage[framemethod=TikZ]{mdframed}
\usepackage{xcolor}
\usepackage{tcolorbox}[boxsep=1mm]
\usepackage{array}

\usepackage[T1]{fontenc}
\usepackage[]{babel}
\usepackage[utf8]{inputenc}

\usepackage{longtable}
\usepackage{lscape}
\usepackage{afterpage}

\mdfdefinestyle{implicationBox}{
    linecolor=blue,
    outerlinewidth=2pt,
    roundcorner=10pt,
    innertopmargin=10pt,
    innerbottommargin=10pt,
    innerrightmargin=10pt,
    innerleftmargin=10pt,
    backgroundcolor=gray!10!white
}

\UseRawInputEncoding

\AtBeginDocument{%
  \providecommand\BibTeX{{%
    \normalfont B\kern-0.5em{\scshape i\kern-0.25em b}\kern-0.8em\TeX}}}

\setcopyright{acmcopyright}
\copyrightyear{2018}
\acmYear{2018}
\acmDOI{XXXXXXX.XXXXXXX}

\acmJournal{TOSEM}
\acmVolume{37}
\acmNumber{4}
\acmArticle{111}
\acmMonth{8}

\begin{document}

%%
%% The "title" command has an optional parameter,
%% allowing the author to define a "short title" to be used in page headers.
\title{Engagement in Code Review: Emotional, Behavioral, and Cognitive Dimensions in Peer vs. LLM Interactions}

%%
%% The "author" command and its associated commands are used to define
%% the authors and their affiliations.
%% Of note is the shared affiliation of the first two authors, and the
%% "authornote" and "authornotemark" commands
%% used to denote shared contribution to the research.

\author{Adam Alami}
\email{adal@cs.aau.dk}
\affiliation{%
  \institution{University of Southern Denmark}
  \streetaddress{The Maersk Mc-Kinney Moller Institute}
  \postcode{6400}  
  \city{S\o nderborg}
  \country{Denmark}
}

\author{Nathan Cassee}
\email{nathancassee@uvic.ca}
\affiliation{%
  \institution{University of Victoria}
  \streetaddress{PO Box 1700 STN CSC}
  \city{Victoria BC V8W 2Y2}
  \country{Canada}
}

\author{Thiago Rocha Silva}
\email{thiago@mmmi.sdu.dk}
\affiliation{%
  \institution{University of Southern Denmark}
  \streetaddress{The Maersk Mc-Kinney Moller Institute}
  \postcode{5230}
  \city{Odense M}
  \country{Denmark}
}

\author{Elda Paja}
\email{elpa@itu.dk}
\affiliation{%
  \institution{IT University of Copenhagen}
  \streetaddress{Rued Langgaards Vej 7}
  \city{2300 Copenhagen S}
  \country{Denmark}
}

\author{Neil A. Ernst}
\email{nernst@uvic.ca}
\affiliation{%
  \institution{University of Victoria}
  \streetaddress{PO Box 1700 STN CSC}
  \city{Victoria BC V8W 2Y2}
  \country{Canada}
  }

%%
%% By default, the full list of authors will be used in the page
%% headers. Often, this list is too long, and will overlap
%% other information printed in the page headers. This command allows
%% the author to define a more concise list
%% of authors' names for this purpose.

\renewcommand{\shortauthors}{Alami, et al.}

%%
%% The abstract is a short summary of the work to be presented in the
%% article.

\begin{abstract}

Code review is a socio-technical practice, yet how software engineers engage in Large Language Model (LLM)-assisted code reviews compared to human peer-led reviews is less understood, especially as artificial intelligence (AI) tools are increasingly integrated into software engineering (SE) workflows. We report a two-phase qualitative study with 20 software engineers to understand such dynamics. In Phase I, participants exchanged peer reviews and were interviewed about their affective responses and engagement decisions. We also prompted them to discuss their submitted code review generated by ChatGPT 4o. In Phase II, we investigated and introduced a new prompt to match engineers' stated preferences for the review and probed how content characteristics shaped their reactions. We develop an integrative account linking \emph{emotional self-regulation} to \emph{behavioral engagement} and resolution. We identify a repertoire of self-regulation strategies that engineers use to regulate their emotions in response to negative feedback: reframing, dialogic regulation, avoidance, and defensiveness. These strategies are inspired by the engineers' commitment to code quality and values: accountability and growth mindset. Engagement proceeds through \emph{social calibration}; engineers align their responses and behaviors to the relational climate and team norms. Trajectories to resolution, in the case of peer-led review, vary by locus (solo/dyad/team) and an internal sense-making process. With the LLM-assisted review, emotional costs and the need for self-regulation seem lower. When LLM feedback aligned with engineers' cognitive expectations (e.g., clear structure, concise scope, neutral tone, actionable), participants reported reduced processing effort and a potentially higher tendency to adopt. Our findings demonstrate that LLM-assisted review redirects engagement from managing emotions and social affect to managing cognitive load. We contribute with an integrative model of engagement, linking \emph{emotional self-regulation} $\leftrightarrow$ \emph{behavioral engagement} $\rightarrow$ \emph{resolution}, showing how affective and cognitive processes influence feedback adoption in peer-led and LLM-assisted code reviews. We conclude that AI is best positioned as a supportive partner to reduce cognitive and emotional load while preserving human accountability and the social meaning of peer review and similar socio-technical activities.

\end{abstract}

%%
%% The code below is generated by the tool at http://dl.acm.org/ccs.cfm.
%% Please copy and paste the code instead of the example below.
%%


%%
%% Keywords. The author(s) should pick words that accurately describe
%% the work being presented. Separate the keywords with commas.

\received{20 February 2024}
\received[revised]{12 March 2025}
\received[accepted]{5 June 2025}

%%
%% This command processes the author and affiliation and title
%% information and builds the first part of the formatted document.
\maketitle

\section{Introduction}
\label{sec:introduction}

\noindent The integration of artificial intelligence (AI) is growing in many processes, including software engineering (SE)~\cite{fan2023large}. However, for decades and despite various practices being automated, SE remains a human activity~\cite{baxter2011socio}. The integration of AI into a socio-technical process such as SE, historically dominated by human control, interactions, and decision-making, may imply a paradigm shift in the way software is engineered.

Research and products aiming to leverage AI for software engineering are targeting different core SE activities. Research efforts have shown interest in requirements engineering, e.g.,~\cite{mich2002nl,ninaus2014intellireq,del2015multi}, software design, e.g.,~\cite{jiao2004automated,rodriguez2016artificial}, construction, e.g.,~\cite{ling2016,yu2018spider}, and testing, e.g.,~\cite{zaeem2014automated,schneid}.

Machine learning (ML) technologies have opened up the possibility to automate complex SE tasks, traditionally entrusted to humans. Advances in neural network architectures, such as recurrent neural networks and transformers~\cite{vaswani2017attention}, have been used to advance the state-of-the-art for many difficult automated software engineering tasks~\cite{dehaerne,fan2023large}. These technologies, among others, have contributed to an array of commercial products such as Tabnine, CodeX, Github's Copilot, and ChatGPT.

While several code-assistant products (e.g., Tabnine, GitHub Copilot, Amazon CodeWhisperer) focus on code-specific tasks such as AI pair programming, general-purpose Large Language Models (LLMs) (e.g., GPT-4-class models) have been used and tested for multiple software engineering activities, e.g., documentation, debugging, design suggestions, and code review~\cite{fan2023large,vaithilingam2022expectation}. The versatility and broad applicability of LLMs have the potential to fundamentally alter the landscape of software engineering by either reducing the reliance on human intervention or support in tasks that were previously considered too complex for automation~\cite{svyatkovskiy2020,zohair2018future}.

Although the tools landscape designated to AI-supported SE and the enthusiasm for their adoption are increasing~\cite{fan2023large,Octovers}, our understanding of the impact of this shift on the social and human dynamics within SE environments remains limited~\cite{anthony2023collaborating}. SE is inherently a socio-technical practice~\cite{baxter2011socio} and its engineering process is entwined with collaboration, human judgment, emotional interactions, and decision-making~\cite{kalliamvakou2017makes}. As we witness a potential paradigm shift, moving from human-dominated processes to AI-supported workflows, it is important to investigate how these changes affect the human and social elements of SE.

Furthermore, most SE activities are collaborative. Thus, the integration of AI into the already complex human and social workflows necessitates a thorough understanding of its implications. Specifically, the introduction of AI in practices traditionally governed by human expertise raises questions about the human-AI engagement and behavioral responses of practitioners like software engineers. For example, Malone et al.~\cite{malone2020artificial} suggest that the most challenging changes in AI integration are not computers replacing humans but rather people and computers working together as integrated ``superminds''. To effectively harness the potential of these ``superminds''~\cite{malone2020artificial}, it is important to understand how software engineers engage with and respond to AI tools in their workflows. To that end, In Phase I of our study, we asked:

\medskip

\noindent \textbf{RQ1:} \emph{How do software engineers perceive and engage with LLM-assisted code reviews compared to their peers?}

\medskip

Engagement, in the context of our study, refers to the ways and the extent to which software engineers actively interact with, respond to, and incorporate feedback from the review process~\cite{oertel2020engagement,doherty2018engagement}. This aligns with the broader definition of engagement as \emph{``the process by which two (or more) participants establish, maintain and end their perceived connection''}~\cite{sidner2004look}, which emphasizes the relational nature of the process. In code review, software engineers establish a temporal ``connection'' with their reviewers (whether peers or LLMs) and sustain it through interactions to reach a resolution to the feedback.
 
We opted for code review as a SE process, because it is inherently a human-centric process that relies on collaboration, judgment, and communication~\cite{badampudi2023modern}. This makes it ideal to evaluate how engineers experience the engagement in an LLM-supported alternative to a process traditionally characterized by human-intensive interactions.

In this Phase I, we set an interview study (20 interviewees) anchored in tangible review experiences. Prior to the interviews, we asked our participants to submit a sample of their own authored code. Then, we assigned two to three human reviewers to every author and asked them to submit written reviews. We used the reviews during the first part of interviews to prompt our interviewees to reflect on their experiences, perceptions, and emotional reactions still salient to them. By using a recent and concrete experience, we grounded our data in a lived-experience context. In the second part of the interview, we shared an LLM-generated review of the participant's authored code to explore how their reactions and engagement differ when interacting with AI-generated feedback as opposed to human-generated feedback. Prior to the interview, all participants were informed with the details of the procedure, i.e., that human- and LLM-generated feedback were to be discussed.

From Phase I analysis, we learned that engagement is multi-dimensional, spanning \emph{cognitive}, \emph{emotional}, and \emph{behavioral} responses. While experiencing these responses, software engineers undertake a \emph{sense-making process} of the feedback in order to make a decision regarding its adoption. This work has been partly published in Alami \& Ernst~\cite{alami2025human}.

While Phase I uncovered these initial insights, it also raised follow-up questions that required deeper understanding. To further unpack the nuances behind these engagement dimensions, such as how engineers move from managing their emotions to resolution of feedback, or how engineers reconcile emotional responses with their professional obligations, we designed Phase II as an in-depth follow-up with some participants (15 out of the 20 participants from Phase I took part). This phase aimed to elicit specific details related to Phase I constructs, and uncover nuances not fully developed in the initial interviews. In this subsequent phase, we specifically sought to understand:

\medskip

\noindent \textbf{RQ2:} \emph{How do software engineers regulate their emotional responses when engaging with peer-generated versus LLM-generated feedback?}

\medskip

\noindent \textbf{RQ3:} \emph{How do software engineers process and act on feedback from peers versus LLMs?}

\medskip

In our prior work~\cite{alami2025human}, we conceptualized the core dimensions of engagement in code review and contrasted peer and LLM-led interactions; however, the behavioral processes and emotional responses that inform engineers' responses to feedback remained underexplored. \textbf{RQ2} and \textbf{RQ3} represent explanatory pathways we sought to explore further in Phase II. While \textbf{RQ2} aims at unpacking the \emph{emotional} aspects of engagement, \textbf{RQ3} dives deeper into the \emph{behavioral} aspects while seeking resolution of the feedback. \textbf{RQ3} also seeks to understand further actionable behaviors that close the engagement loop. The separation of \textbf{RQ2} and \textbf{RQ3} is intended to find nuanced insights we did not observe in our previous work~\cite{alami2025human} (presented in this study as \textbf{RQ1} for context). We could not find nuanced insights in \textbf{RQ1} because it is foundational, focusing on identifying and characterizing the dimensions of engagement, rather than the emotions and behaviors that unfold while engineers seek resolution. 

Phase I also revealed a spectrum of reactions and preferences toward LLM feedback, potentially influencing the likelihood of adoption, specifically the high cognitive load required to process the LLM's feedback. This variation across participants warranted further exploration. Hence, we asked:

\medskip

\noindent \textbf{RQ4:} \emph{How do the content characteristics of LLM-generated feedback (e.g., scope, format, clarity, tone) shape software engineers' engagement and adoption in LLM-assisted code review?}

\medskip

In Phase I, we used a baseline prompt to generate the reviews. This technique showed structural inconsistency in the outputs. We also learned in the interviews that the reviews require more mental effort in processing the content. Given the minimal control over the output we experienced in Phase I and to evaluate \textbf{RQ4} with reviews aligned to participants' preferences, in Phase II, we investigated the prompt techniques to enhance output alignment (see Sect.~\ref{sec:methods}).

These RQs maintain conceptual alignment with Phase I, while seeking depth, and behavioral specificity, all of which are in line with follow-up qualitative design~\cite{miles2014qualitative}. In Phase II, we conducted 15 follow-up interviews with the same participants of Phase I. In this manuscript, Phase I is reported as \textbf{RQ1}, using a baseline prompt. Phase II encompasses \textbf{RQ2--4}. Collectively, our findings (Phase I and II) show:

\begin{itemize}

   \item [-] \textbf{Multi-dimensional engagement in code reviews:} We identified the intricate dimensions of engagement in code review, which include cognitive, emotional, and behavioral responses to feedback. We also identified how the introduction of LLMs could influence and shape these engagement attributes. For example, our study shows that LLM-assisted review reorients engagement from managing emotional and social affects to managing cognitive load, correctness of LLM's comments, and context alignment.

   \item [-] \textbf{Human-AI collaboration in code review:} Our study contributes to furthering our under\-standing of how software engineers perceive and are willing to collaborate with AI tools. These insights highlight future implications of introducing AI technologies in a socio-technical process like code review. For example, our study identifies the construct of \emph{feedback alignment}, i.e., how LLM feedback structure and tone shape cognitive ease and adoption likelihood.

   \item [-] \textbf{A model of engagement in code review:} Our study also contributes conceptually to the understanding of software engineering as a socio-technical practice. We introduce a conceptual model, linking \emph{emotional self-regulation} $\rightarrow$ \emph{behavioral engagement} $\rightarrow$ \emph{resolution}, showing how affective and cognitive behaviors may influence feedback uptake. We also identifies future implications on the model, as AI is increasingly integrated into SE workflows and practices (see Sect.~\ref{sec:discussion}).
    
\end{itemize}

In the remainder of this paper, we review related work in Sect.~\ref{sec:related}, the key findings or our earlier work~\cite{alami2025human} in Sect.~\ref{sec:background}, and describe our methods in Sect.~\ref{sec:methods}. Then, we report our findings in Sect.~\ref{sec:findings}, discuss their implications in Sect.~\ref{sec:discussion} and trustworthiness in Sect.~\ref{sec:trust}, and conclude the paper in Sect.~\ref{sec:conclusion}.

\section{Related Work}
\label{sec:related}

While code review is part of our research context, it is not the primary focus of our investigation. We used code review as a lens through which we evaluated a broader and critical issue of how software engineers perceive and engage with AI tools in a socio-technical process. In particular, we examine how engineers respond to AI-generated feedback compared to feedback from peers. Therefore, the scope of our related work is to draw on studies focusing on human-AI collaboration and how software engineers engage with AI in SE processes. We set our results in the context of existing literature, including code reviews, in a discussion section (see Sect.~\ref{sec:discussion}).

Human-tool integration before the advent of AI (i.e., LLM-based assistants) often took the form of reporting automated build and test results or other static analysis checks (e.g., linting or security scans). A simple example is the output of a compilation run and the resulting error messages. Even this relatively simple set of tool results can be difficult for humans to understand and work with~\cite{Denny2021}. Static analysis results might take the form of reporting that a particular code file contains known security issues (e.g., possible SQL injections). This is a considerably more actionable and explicable report than the detailed AI-generated code reviews we consider in this paper. But even these simpler reports are challenging to integrate into human decision-making. For example, Johnson et al.~\cite{johnson2013} showed that reporting the factual outputs of the analysis is not sufficient for developers to adopt a static analysis tool.

A common medium for human-tool integration is a bot, typically serving as a textual/chat interface to the underlying static analyzer. For example, Repairnator was a bot for GitHub that was an interface to a sophisticated program repair tool~\cite{Monperrus2019}. The bot's designers found that without \emph{explanation} it was hard or impossible to get humans to accept the changes~\cite{Monperrus2019b}.

In a wider study, researchers examined bot trust and autonomy and found that individuals have varying levels of trust in the bot's recommendations~\cite{ghorbani2023autonomy}. That paper recommended that bot interaction characteristics should be customizable. In addition, \emph{how the bot appears} (i.e., avatar, language, name) can have a big impact on human acceptance: merely revealing that something is a bot can dramatically change human perceptions~\cite{Murgia2016}.

These perceptions may be changing as LLM capabilities make AI assistance more commonplace. Nonetheless, Al Haque et al.~\cite{alhaque} have identified several challenges, including that the AI-generated content was found to be overly polite, overly detailed, and lacked trustworthiness. To date, most of the work on LLMs and coding, including code review, has focused on the technical aspects of LLMs rather than human interaction issues. Technical work looked at creating datasets of pull requests and code problems~\cite{Li2022}, examining how to improve fine-tuning, e.g., LLaMA-Reviewer~\cite{Lu2023}, and exploring different architectures for the underlying neural networks~\cite{Tufano2022}.

Lack of specific context is a challenge for AI tools. Developers get annoyed when the AI does not understand the locality of their codebase, although for boilerplate or generic code, developers appreciate saving keystrokes~\cite{Liang2024}. As Panichella et al.~\cite{Panichella2020} point out, since code review tasks themselves are frequently changing, e.g., as the organization develops a new understanding of concurrency, statically trained LLMs may miss evolving changes.

In sum, these studies show that achieving effective human-tool integration is about much more than functionality~\cite{Li2022,Lu2023,alhaque}. Tool design must also address human aspects such as trust, interpretability, and engineers' preferences for communication style and content~\cite{ghorbani2023autonomy,Murgia2016}. Our study contributes to this body of work by identifying potential shifts in software engineers engagement in response to AI-driven code review.

\section{Background: Earlier Work}
\label{sec:background}

\begin{figure*}[th!]

\includegraphics*[trim=0.5cm 7.1cm 1.5cm 1cm, clip, width=.90\textwidth]{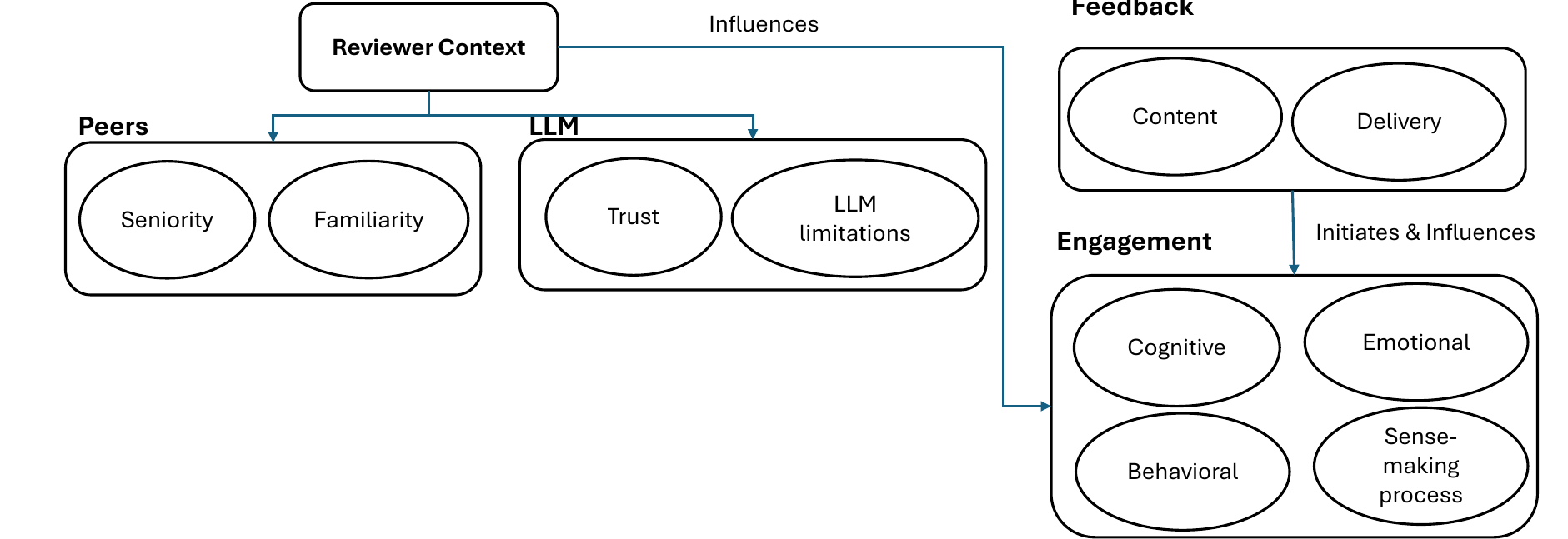}

\caption{An abstract presentation of the findings from Alami \& Ernst~\cite{alami2025human}, capturing engagement and how engineers respond to feedback. The line from \textbf{Reviewer Context} to \textbf{Peer} and \textbf{LLM} indicates two type of reviewers, and the ovals inside the rectangles are elements of the reviewer contexts.}

\label{fig:fig1}

\end{figure*}

In our earlier study~\cite{alami2025human}, we examined how software engineers perceive and engage with LLM-assisted code reviews compared to their peers. We used semi-structured interviews with 20 software engineers. We grounded the data collection in the participants' lived experiences; i.e., all participants submitted their own authored code and reviewed other participants' code. This approach allowed us to anchor the interviews in a recent lived and concrete experience. During the interviews, we asked participants to reflect on both human- and LLM-generated feedback on their submitted code.

Our analysis showed that engagement in code review is multi-dimensional, encompassing \textbf{cognitive}, \textbf{emotional}, and \textbf{behavioral} dimensions. Engineers often undertook a \textbf{sense-making process} to evaluate the feedback before deciding whether to act on it, irrespective of whether the feedback originated from peers or LLMs.

The contribution of this initial investigation was a conceptual model of engagement in code review (see Fig.~\ref{fig:fig1}). In the case of peers, the model illustrates how engagement is influenced by both the content and delivery style of feedback (e.g., content and tone), as well as the reviewer context, which includes familiarity between the reviewer and the author, and perceived expertise of the reviewer.

Human reviews triggered a wider spectrum of emotional reactions, either negative or positive, depending on tone and perception of the content. In contrast, the engagement with LLM feedback often reduced emotional taxation due to the consistently polite tone of the LLM. However, LLMs introduced higher cognitive load due to extensive and sometimes less contextualized feedback.

These findings laid the groundwork for the current study, which seeks to deepen our understanding of the behavioral mechanisms underlying these engagement dynamics. Building on this foundation, the present study extends our earlier work in three ways. First, it moves from an abstract explanation of engagement to examining how its dimensions, emotional, and behavioral processes, unfold in both peer-led and LLM-assisted reviews. Second, in this extension, we introduce the full pathway, from emotional response to behavioral engagement and resolution. Third, we investigated an improved prompt design in Phase II to test how tailoring LLM feedback to engineers' preferences may influence their cognitive effort, perceived alignment, and likelihood of adoption. Collectively, these extensions improve the abstract model of engagement from our previous work~\cite{alami2025human} into a process-oriented and detailed model that captures the dynamics of how engineers regulate, interpret, and act on feedback in increasingly hybrid human-AI SE.
\section{Methods}
\label{sec:methods}

We opted for a 2-phased interview study to investigate our \textbf{RQs}. This choice is better suited when the phenomenon under study requires uncovering complex social and behavioral nuances that influence human judgment and interactions in socio-technical processes~\cite{kvale2009interviews,baxter2011socio}. Interviews are also a good tool to access perspectives and insights ingrained deeply in personal beliefs, attitudes, and behaviors~\cite{larkin2021interpretative,seidman2006interviewing}. By tapping into these core personal and deeply-seated cognitive constructs~\cite{larkin2021interpretative,seidman2006interviewing}, we aim to collect data aligned with our study objectives.

Phase I served as a foundational exploration, with substantial depth, and Phase II as an elaborative follow-up of our initial findings. In Phase I, we identified the core constructs of engagement and the interplay between feedback source (LLM vs. peer) and engineers' engagement and responses. These findings also revealed the need for nuanced understanding of the engagement with the LLM, mainly fine-tune our LLM prompt engineering strategy. Accordingly, in Phase II, we carried out follow-up interviews to collect more data and conduct further analysis to (1) examine how engineers rationalize their behavioral responses to LLM feedback when aligned with their preferences, (2) revisit key constructs such as cognitive load and sense-making in more depth in LLM-assisted review, and (3) fine-tune our LLM prompt engineering strategy to accommodate the divergent preference for feedback style and format we learned in Phase I. The prompt engineering experiment allowed us to align the LLM feedback with the participant's preferences and collect additional data to compare with Phase I.

In the remaining of this section, we describe our research process, including both phases, and methodological decisions and their implementations. We commence in next subsection (Sect.~\ref{sec:rec}) by detailing the recruitment process of Phase I. In subsection~\ref{sec:data}, we explain the data collection process and methods we used in Phase I. Then, in subsection~\ref{sec:phase_2}, we outline the follow-up interview process of Phase II. We document our prompt engineering experiment of Phase II in subsection~\ref{sec:prompt} and the data analysis of both phases in subsection~\ref{sec:analysis}. Section~\ref{sec:trust} is dedicated to our trustworthiness methods. 

\subsection{Phase I: Interviewees recruitment \& selection}
\label{sec:rec}

\begin{table*}[t!]

  \begin{center}
    \footnotesize
    \caption{Interviewees' characteristics. ``NB'' refers to non-binary/third gender. Asterisk (*) indicates those who participated in the follow-up interviews of Phase II.}
    
    \label{tbl:population}
    \renewcommand\arraystretch{0.70}
        
    \begin{tabularx}{\textwidth}{l|l|l|l|X|l|l|l}

     \toprule
      \textbf{\#} & \textbf{Role} & \textbf{Exp.} & \textbf{Gender} & \textbf{Industry sector} & \textbf{Language} & \textbf{Reviewers} & \textbf{Country}\\
     \midrule
    
        P1 (*)& Sr. Soft. Eng. & 6-10 years & Male & IT Services & JavaScript & P10, P19, \& P20 & Germany\\
        P2 (*)& Sr. Soft. Eng. & 6-10 years & Male & Consulting & Java & P1, P3, \& P19 & UK\\
        P3 (*)& Sr. Soft. Eng. & >10 years & Male & Telecom & Python & P4, P12, \& P18 & UK\\
        P4 (*)& Software Eng. & 3-5 years & Male & Telecom & Python & P3, P12, \& P20 & Germany\\
        P5 & Software Eng. & >10 years & Male & IT Services & C\# & P11 \& P17 & USA\\
        P6 (*)& Software Eng. & 6-10 years & NB & Finance & Java & P6, P11, \& P14 & Netherlands\\
        P7 (*)& Tech Lead & >10 years & Male & Finance & Java & P11 \& P14 & Canada\\
        P8 (*)& Software Eng. & 3-5 years & Male & Aviation & Python & P9, P17, \& P20 & Spain\\
        P9 & DevOps Eng. & >10 years & Female & Health Care & Python & P8, P13, \& P20 & UK\\
        P10 (*)& Software Eng. & 1-2 years & Male & IT Services & Python & P3, P12, \& P18 & UK\\
        P11 & Software Eng. & 3-5 years & Female & Finance & Java & P6 \& P7 & South Africa\\
        P12 (*)& Sr. Soft. Eng. & >10 years & Female & Finance & Python & P4, P12, \& P19 & UK\\
        P13 (*)& Sr. Soft. Eng. & 6-10 years & Male & IT Services & TypeScript & P16 \& P17 & Ireland\\
        P14 (*)& Software Eng. & 6-10 years & Male & IT Services & Java & P5, P8, \& P13 & Portugal\\
        P15 & Sr. Soft. Eng. & 3-5 years & Female & Government Serv. & Python & P16 \& P18 & Italy\\
        P16 (*)& Software Eng. & 1-2 years & Male & IT Services & JavaScript & P6, 13, \& P21 & Portugal\\
        P17 (*)& Software Eng. & 3-5 years & Male & Media \& Entert. & Python & P5, P8, \& P9 & UK\\
        P18 & Software Eng. & 6-10 years & Male & IT Services & C++ & P2, P12, \& P15 & UK\\
        P19 (*)& Sr. Soft. Eng. & 6-10 years & Female & IT Services & Java & P2 \& P10 & Germany\\
        P20 (*)& Software Eng. & 3-5 years & NB & IT Services & JavaScript & P1 \& P10 & Ireland\\
    
     \bottomrule
     
    \end{tabularx}
   
  \end{center}
  
\end{table*}

To recruit our interviewees, we used Prolific\footnote{\url{https://www.prolific.co/}}, a research market platform. Given that the platform does not verify nor evaluate self-reported skills~\cite{alami2024you}, we carried out a pre-screening process to qualify our potential participants, following the guidelines suggested by Alami et al.~\cite{alami2024you}. 

Alami et al.~\cite{alami2024you} recommend using an iterative and controlled prescreening process, using task-oriented questions that go beyond theoretical understanding to help filter genuine engineers from those who may rely on external resources, such as Googling or prompting an LLM for answers. 

In the pre-screening questionnaire, we used a programming task, a critical-thinking question, and a question for participants to share a problem-solving scenario from their own experience. We iteratively and manually assessed the answers~\cite{alami2024you}. First, we evaluated the answers for AI-generated content using ChatGPT 4o. Subsequently, we manually evaluated the content's quality and coherence~\cite{alami2024you}. In this initial pre-screening, we capped the number of participants to 500. We iteratively assessed the pre-screening questionnaire answers as they were submitted. After the assessment, we ended up with 353 qualified engineers.

Then, we ran a second and final pre-selection questionnaire, where we asked for further demographic data, programming language skills, and whether the participants were willing to participate in an interview study. We limited this pre-selection to 100, and 76 agreed to participate in the interviews.

This two-phased approach had different objectives. While in the initial prescreening, we focused on evaluating potential participants' skills, i.e., whether they were genuine software engineers~\cite{alami2024you}, in the second pre-selection, we wanted to know whether those qualified software engineers were willing to take part in an interview study, collecting further demographic data and the programming languages they may submit their code and review other participants' codes. We paid \textsterling0,50 for the initial prescreening and pre-selection and \textsterling60,00 (up to 60 minutes) for the participation in the first interview, including the time spent to review other participants' code and \textsterling30,00 for the second phase follow-up interviews (30 to 40 minutes) (Phase II). The selection process took place in August 2024. The questionnaire used in the pre-selection is available on the replication package (see Sect.~\ref{sec:replication}).

Table~\ref{tbl:population} documents our participants' characteristics, providing their current roles, experience levels, gender, industry sectors, and the programming languages they opted to submit their code in. The column ``Reviewers'' lists the reviewers assigned to each participant for the peer-led reviews. We managed to assign to each participant a minimum of two reviews. Even though we aimed for three reviews per participant to align with industry and open-source community practices~\cite{alami2020foss,alami2021pull}, we did not manage to meet this expectation due to the varying programming skills and language preferences among participants, which limited the pool of suitable reviewers for certain code submissions. In assigning reviewers to authors, we aimed for a diverse range of experiences and backgrounds. For example, P1, a senior software engineer with an experience of 6--10 years, was assigned reviewers with experiences ranging from less than two years to ten years. This diversity in reviewers' experience levels was intentionally designed to enrich our data with varying levels of expertise and perspectives and feedback from both novice and seasoned reviewers. We also aimed to be diverse in gender and country in the reviewers selection. For example, two females and one male from two different countries reviewed P18.

\subsection{Phase I: Data collection}
\label{sec:data} 

We designed a research process that mirrors similar professional practices, yet within the constraints inherent to a research environment. For example, our participants remained anonymous to each other, opposed to a professional context where authors and reviewers are known to each other. For Phase I, upon the completion of our recruitment and selection process, we asked our participants to submit a code they had authored. Then, we assigned to each author two to three reviewers (drawn from other participants in the study, see Table~\ref{tbl:population}) and asked them to submit a written review as if they were to conduct a peer review in their professional contexts. Prior to the interviews, we shared the peer reviews with the authors. Part of the recruitment process, we asked the participants permission to prompt ChatGPT 4o to review their submitted code and informed them that we would share the LLM-generated reviews in the interview for discussion.

\begin{figure*}[!t]
    \includegraphics*[trim=0cm 5.6cm 0cm 1.5cm, clip, width=1.0\textwidth]{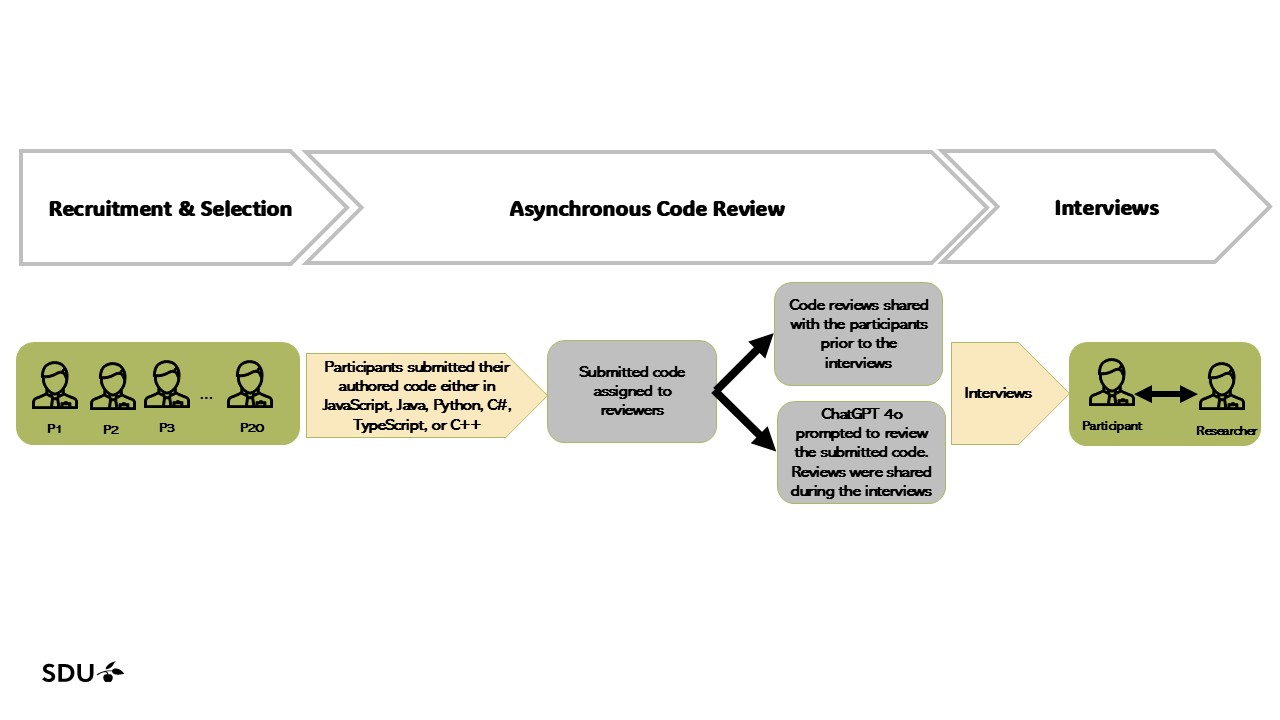}
        \caption{Phase I Research Process.}
        \label{fig:Phase1}
        
\end{figure*} 

In the first part of the interview, we used the reviews (feedback received from other participants on the interviewee's code) as catalysts to prompt the engineers to explain and elaborate further on their responses to our questions. In the second part, we shared an LLM-generated review of the same code authored by the interviewees to prompt them to share their attitudes and understand how they would engage with it compared to human-authored reviews. This approach ensures that we capture the nuanced differences in how software engineers engage with and respond to both human and LLM-generated feedback. First, we grounded the discussion in their real-world practices, assigned task, and the study's review experience. Then, we prompted them to contrast their earlier claims when reviews are LLM-generated. This dual-method strategy, combining data in both engagement with peer-conducted and LLM-generated reviews, aligns with the study objectives. The process was explained to participants during the recruitment in details prior to the interview. Figure~\ref{fig:Phase1} summarizes Phase I process.

We opted for ChatGPT for its wide accessibility and familiarity among a diverse audience. In Phase I, to generate ChatGPT reviews, we used this prompt: \emph{``You are an expert of [the programming language]. Provide a thorough review of the attached code.''} This prompt design aimed to generate expert-level feedback, relatable to what a human reviewer might offer~\cite{brown2020language}. Such feedback is what developers would expect from knowledgeable peers~\cite{brown2020language,liu2023pre}.

Our interviews were semi-structured. Although we designed an interview guide, its purpose was to guide the conversation while allowing fluidity by prompting the interviewee to elaborate and explain further based on their responses. We also used examples from the reviews they received to engage them in deeper reflection of how they perceived the comments they received either from the LLM or other participants. Table~\ref{tbl:guide} illustrates examples of the questions we asked. The full guide is available in the shared documents package (Sect.~\ref{sec:replication}).

All interviews were conducted using Zoom and lasted 40-60 minutes, with a total of 17h12min of audio. After transcription, the audio generated a total of 271 pages verbatim, an average of approximately 14 pages per interview. The first author conducted the interviews in the first and second weeks of September 2024. We used Otter.ai\footnote{\url{https://otter.ai/}}, an online transcription tool, to transcribe the audio recordings.

\begin{table}[th!]
  \footnotesize
  \caption{Key parts of the interview guide (Phase I)}%

  \renewcommand\arraystretch{1.3}

  \begin{tabular}{p{13.5cm}}
     \hline
     \textbf{Introduction}\\
     \midrule
    
      Can you please introduce yourself? (Include educational background, current role, and years of experience in software development.)\\
      
      \hline
      \textbf{Section I: Approach to reviewing code}
      \\\midrule
 
        Are there any coding standards or best practices that you kept in mind during the reviews we assigned to you?\\
        What did you consider most when you wrote your feedback for the assigned code?\\
    
     \hline
     \textbf{Section II: Engagement with peers' feedback on authored code}
     \\\midrule

        How do you typically feel when you receive feedback on your code from peers?\\
        In the feedback you received from the other participants, what elements of the feedback do you find most useful or valuable? And why?\\

     \hline
     \textbf{Section III: Engagement with LLM-generated review}
     \\\midrule

        How did you feel when you first read the LLM-generated review of your code?\\
        Would you respond and react to the LLM-generated feedback the same way as you do to human's feedback?\\

    \hline
     \textbf{Conclusion}
     \\\midrule

        Based on your experiences in this study, how would you summarize the main differences between peer and LLM-generated code reviews?\\
        Is there anything else you would like to share about your experience in this study or with code reviews in general?\\

    \bottomrule
    
  \end{tabular}
  
  \label{tbl:guide}
    
\end{table}

\subsection{Phase II: Follow-up Interviews}
\label{sec:phase_2}

Upon the completion of Phase I data analysis (see Sect.~\ref{sec:analysis}), we decided to conduct follow-up interviews with Phase I participants. While Phase I provided initial, yet interesting insights, in this follow-up phase, we sought additional data to elaborate the empirical support for \textbf{RQ4} and enrich \textbf{RQ1--RQ3} with additional evidence.

\begin{figure*}[!t]
    \includegraphics*[trim=0cm 3.5cm 0cm 0.5cm, clip, width=1.0\textwidth]{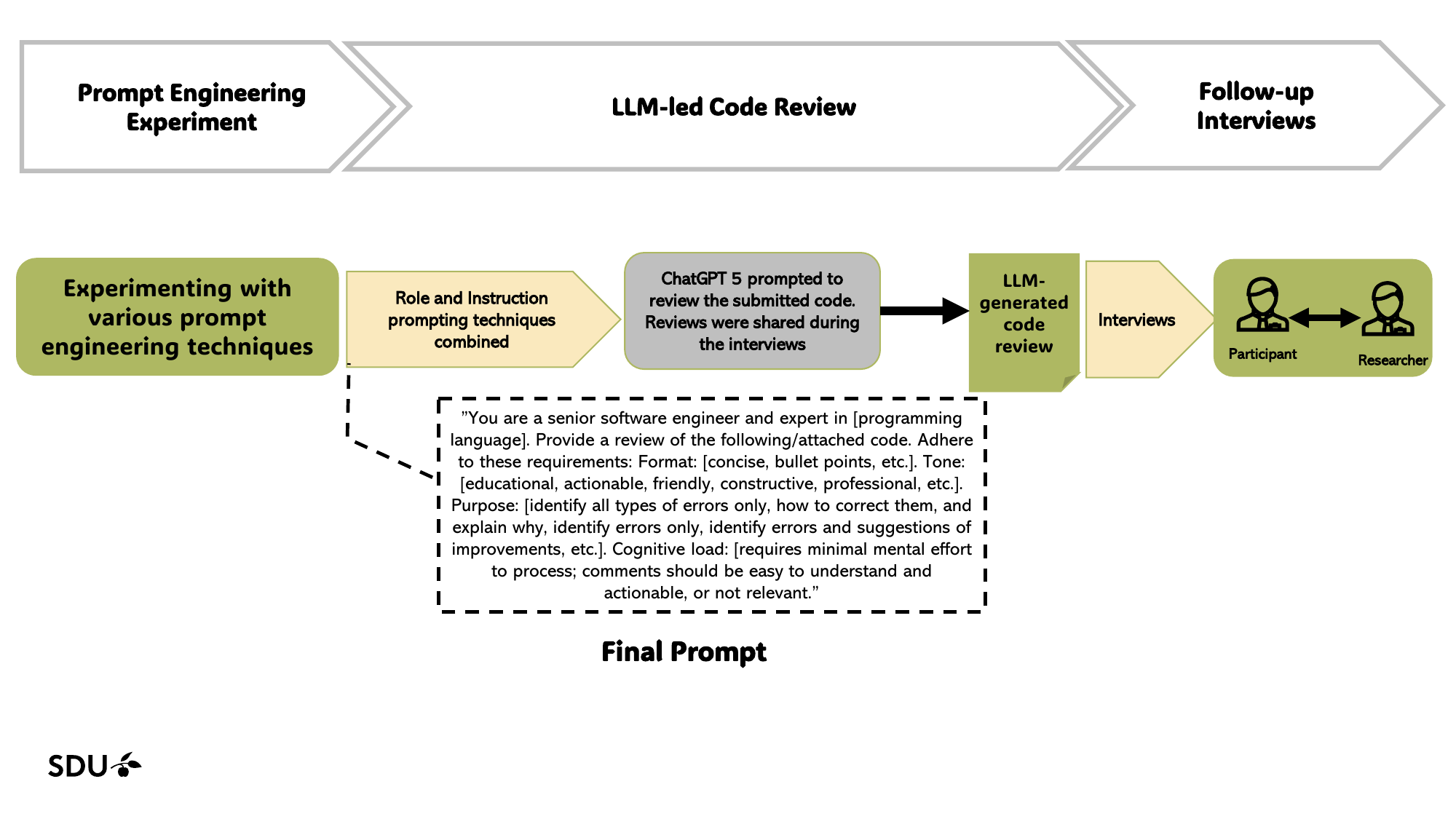}
        \caption{Phase II Research Process.}
        \label{fig:Phase2}
        
\end{figure*}

Figure~\ref{fig:Phase2} summarizes the process we followed in Phase II, which integrates a prompt engineering experiment with follow-up interviews. The figure illustrates how we first experimented with individual and combined prompt engineering techniques, ultimately selecting a combined ``Role and Instruction'' prompting strategy to generate LLM reviews aligned with participants' preferences. These individual tailored reviews were then shared during follow-up interviews (i.e. each participant's code had a new individually tailored review aligned with their preferences), enabling us to explore how alignment in feedback format, tone, and purpose may influence cognitive load, sense-making, and behavioral responses compared to Phase I.

Given that the intent of Phase II is to elaborate and expand on constructs identified in Phase I and collect new data for \textbf{RQ4}, we adopted a confirmatory yet exploratory structure. We organized the guide using a layered approach, aligned with \textbf{RQ2--RQ4} while remaining open to new insights, using prompting questions. The design of the interview guide followed these principles: \textbf{(1) Thematic anchoring:} We aligned the structure of the guide with our RQs (cognitive load and sense-making, emotional regulation, behavioral responses)~\cite{kvale2009interviews,smith2021interpretative}. \textbf{(2) Hypothesis probing:} Drawing from our Phase I themes, we constructed hypothesis-informed questions, while avoiding leading formulations~\cite{kvale2009interviews,smith2021interpretative}. \textbf{(3) Personalized grounding:} We incorporated claims from the participant's own Phase I transcript and feedback received from other participants and the LLM, to stimulate richer elaboration~\cite{kvale2009interviews,smith2021interpretative}. \textbf{(4) Contrast \& reflection:} We used reflective prompts throughout the interview to elicit data to compare past and current experiences, and whether any changes in perceptions occurred since Phase I~\cite{kvale2009interviews,miles2014qualitative,saldana2021coding}.

Accordingly, we structured the guide around four sections: (1) emotional regulation (\textbf{RQ2}), (2) behavioral engagement \textbf{(RQ3)}, (3) sense-making and cognitive load in LLM-assisted reviews (\textbf{RQ4)}, and (4) reflective reappraisal. Table~\ref{tbl:followupguide} summarizes the structure and examples of our Phase II questions.
 
We contacted all Phase I participants and invited them to take part in the follow-up phase. Fifteen of them accepted ((*) in Table~\ref{tbl:population} indicates Phase II participants). The follow-up interviews took on average 30 minutes each and generated seven hours and forty minutes of audio. The audio generated a total of 121 pages verbatim after transcribing, an average of approximately eight pages per follow-up interview. The first author conducted the interviews in August 2025.

\begin{table}[th!]

  \footnotesize
  
  \caption{Key parts of the follow-up interview guide (Phase II)}%
  
  \renewcommand\arraystretch{1.3}
  
  \begin{tabular}{p{13.5cm}}
     \hline
     
     \textbf{Introduction \& shifts in perception since Phase I}\\
     \midrule
     Reminder of prior participation and purpose of follow-up.\\
     Participant was invited to reflect on prior interview experience and any updates.\\
     Have you integrated AI in your workflow since our last conversation? In what ways?\\
     \hline
     
     \textbf{Section I: Emotional self-regulation (RQ2)}\\
     \midrule
     You previously mentioned [we used quote from Phase I indicating emotion]. Can you tell me more about how you typically handle such reactions during reviews from peers?\\
     Do you recall any recent experiences, peer or LLM, where you felt the need to manage your emotions? And how do you manage your emotions in the case of negative feedback?\\
     What helps you respond constructively when feedback feels off or hard to accept?\\
     \hline
     
     \textbf{Section II: Behavioral engagement (RQ3)}\\
     \midrule
     Can you describe what you usually do after receiving feedback from a peer? And what would you do if we replace your peer with an LLM like ChatGPT?\\
     When you received feedback from the LLM and other participants in this study, did you act on one type of feedback more readily than the other? Why?\\
     \hline
     
     \textbf{Section III: Cognitive load and sense-making of LLM review (RQ4)}\\
     \midrule
     How did you go about making sense of the feedback from ChatGPT compared to your peers?\\
     Were there instances where the LLM feedback felt harder or easier to interpret? Can you walk me through one example?\\
     With this new generated feedback from the LLM, have your perceptions of the usefulness or clarity of AI feedback changed?\\
     \hline
     
     \textbf{Section IV: Shifts in perception since Phase I}\\
     \midrule
     Have your views on AI in code review changed since the first interview?\\
     Would you prefer to receive more LLM-generated feedback, less, or about the same? Why?\\
     What do you now believe makes feedback --- AI or human --- actionable or meaningful?\\
     \hline
     
     \textbf{Conclusion} \\
     \midrule
     Would you like to share anything we have not covered in the interview?\\
     Reconfirm consent for data and transcript use.\\
    \bottomrule
    
  \end{tabular}
  
  \label{tbl:followupguide}
  
\end{table}

\subsection{Phase II: Prompt Engineering}
\label{sec:prompt}

In Phase I, we learned that the engineers in our sample had varying expectations regarding the format, style, and tone of LLM-generated reviews. Content-related features, such as verbosity, generic positivity, and over-explanation in the LLM's reviews, led to disengagement.

To better align the LLM's feedback with engineers' expectations, to address \textbf{RQ4}, and provide a contrast with Phase I, we reviewed the literature on best practices in prompt engineering~\cite{chen2023unleashing,WhatisPr}. Then, we carried out an evaluation of selected prompt engineering techniques to evaluate their efficiency in generating the most relevant outputs for the follow-up phase.

Analytically, the purpose of this prompt engineering experiment is to identify the best techniques to prompt the LLM to generate feedback aligned with our participants' individual preferences. More specifically, to examine variation in the concepts we identified in Phase I, mainly the relationship between the cognitive load required to process the feedback and the sense-making process that follows when the feedback is aligned with the engineers' preferences for format, tone, and the purpose of the review.

To systematically evaluate and compare prompt engineering techniques in generating LLM-based code reviews that align with the diverse preferences identified in Phase I, we evaluated a few prompt engineering techniques, namely Zero-Shot~\cite{logan2022cutting}, Instruction~\cite{mishra2021reframing}, and Role prompting~\cite{zhang2023visar,chen2023unleashing,zheng2024helpful}. Each technique was evaluated in isolation as well as in combination with the others. We documented our prompt engineering experiment in an audit trail report, available in our replication package (see Sect.~\ref{sec:replication}). The link to ChatGPT chat we used for both experimentation and generating the reviews of Phase II can be found \href{https://chatgpt.com/share/6890abe6-5af0-800c-a8ae-e065d90e1434}{here}\footnote{\url{https://chatgpt.com/share/6890abe6-5af0-800c-a8ae-e065d90e1434}}.

We intentionally limited our experiment to prompting techniques that demonstrated performance across broad task types. Based on our review of prompt engineering literature~\cite{chen2023unleashing,WhatisPr} and Phase I findings, we focused on achieving participant-aligned feedback, seeking a clear link between prompt structure and output variation. This focus favored Zero-Shot, Instruction, and Role prompting as first candidates. Other techniques, such as Retrieval-Augmented Generation, Tree-of-Thought prompting, or AutoPrompt, were not considered at first because they are less controllable for variation like tone and style alignment~\cite{wei2022chain}. After testing our initially selected prompts (i.e., Zero-Shot, Instruction, and Role prompting), we did not deem pursuing other strategies warranted; our selected strategy, Instruction $+$ Role, has yielded superior results.

To carry out the experiment, we selected examples from the code submitted in Phase I. We selected a representative sample of six code snippets, ensuring variation in programming language (e.g., Java, Python, JavaScript), complexity, and domain (e.g., finance, logic-heavy, algorithmic). This diversity reflects the variation present in Phase I code dataset. This variety also strategically sought to identify any changes in feedback style that may surface in the LLM outputs. Table~\ref{tbl:prompt_code_selection} summarizes selected code for the experiment and the selection rationale.

\begin{table*}[ht!]

  \footnotesize
  
  \caption{Selected code samples for prompt engineering experiment}
  \renewcommand\arraystretch{1.4}
  
  \begin{tabular}{l p{3.2cm} c p{1.2cm} p{6cm}}
    \toprule
    
    \textbf{P\#} & \textbf{Domain} & \textbf{Complexity} & \textbf{Language} & \textbf{Rationale for Selection}\\
    \midrule
    
    P1 & Version Control Utility & Medium & JavaScript & Illustrates CLI (command line interface) interaction and semantic versioning logic; suitable for evaluating clarity and conciseness in feedback.\\
    
    P2 & Banking / Financial Services & High & Java & Contains business logic, custom exceptions, and unit tests; useful for testing review purpose alignment (educational vs. corrective).\\

    P5 & 3D Rendering / Game Engine & High & C\# & Includes graphics buffer handling and object-oriented complexity; selected to test LLM tone and completeness in abstract domains.\\

    P12 & Mathematical Algorithms & Low & Python & Basic algorithmic function; serves as a baseline for evaluating verbosity and tone in LLM-generated feedback.\\

    P13 & Cloud Function Routing & Medium & TypeScript & Decision logic for backend lambda function execution; suitable for testing clarity and error-handling feedback.\\

    \bottomrule
    
  \end{tabular}
  
  \label{tbl:prompt_code_selection}
  
\end{table*}

We defined four criteria inspired by Phase I findings on LLM feedback usability. We used these criteria as requirements on the prompts to instruct the LLM on format, style, and tone of the output. For example, we identified ``format'' as a variable in the prompt, because we wanted to influence the LLM format when the participant has preference for a specific format, like bullet points. Similarly, ``actionability'' was identified to instruct the level of specificity required in the feedback. Table~\ref{tbl:prompt_eval_criteria} documents the criteria we used in the experiment.

\begin{table}[ht!]

  \footnotesize
  \caption{Evaluation criteria for prompt engineering outputs}
  \renewcommand\arraystretch{1.4}
  
  \begin{tabular}{p{3cm} p{10cm}}
    \toprule
    
    \textbf{Criterion} & \textbf{Description}\\
    \midrule
    
    \textbf{Format} & This criterion is to instruct the LLM the structure and the format of the feedback.\\
    
    \textbf{Actionability} & Does the review provide specific, implementable suggestions that the developer can act on immediately?\\
    
    \textbf{Tone Alignment} & Does the tone of the review reflect the participant's preferences (e.g., polite, direct, collaborative)? We used this criterion to instruct the LLM to use a specific tone in phrasing the feedback.\\
    
    \textbf{Cognitive Load} & Is the review concise enough to reduce unnecessary mental effort while still delivering meaningful content? We used this variable to indicate to the LLM that it must cater for this requirement when the engineer prefers low mental load in processing the feedback.\\
    
    \textbf{Purpose Fit} & Does the review serve the participant's intended purpose (e.g., contextualized feedback, coding improvement, educational guidance)? We used this variable to instruct the LLM the engineer's preference for the purpose of the feedback. For example, in Phase I, some of our participants wanted straightforward feedback to improve code; others wanted also a feedback that they can learn from.\\
    
    \bottomrule
    
  \end{tabular}
  
  \label{tbl:prompt_eval_criteria}
  
\end{table}

We tested every technique in isolation sequentially; Zero-Shot, followed by Instruction, then Role prompting. In the final tests and based on earlier results, we combined Instruction and Role prompting. This sequence mirrors a layered prompt design strategy~\cite{mishra2021reframing}. We started with minimal scaffolding and progressively added structure and control to the prompt~\cite{mishra2021reframing,mishra2025promptaid}. This strategy allowed us to observe the incremental effect of each technique and control for confounding interactions. A confounding interaction occurs when two (or more) prompt techniques are used together and it becomes unclear which one caused the observed effect~\cite{mishra2021reframing}.

By testing each technique in isolation, we managed to understand better its unique contribution to review quality and its ability to align with our criteria (Table~\ref{tbl:prompt_eval_criteria}), before exploring combinations (e.g., Instruction $+$ Role). This modular comparison provided insights into how each prompt strategy affects our evaluation criteria (format, actionability, purpose fit, and tone), allowing findings from early stages of testing to inform adjustments or combinations in later stages. Our approach also aligns with established literature. For example, Mishra et al.~\cite{mishra2021reframing} and Wei et al.~\cite{wei2022chain} recommend starting from basic prompting to isolate effects before layering techniques.

We found that combining instruction and role techniques yields superior results. Upon the completion of the evaluation of prompt techniques for \textbf{RQ4}, we re-generated ChatGPT 4o reviews for the code submitted in Phase I. The newly generated reviews were used in the follow-up interviews. Our final prompt is:

\medskip

\begin{center}

    \emph{You are a senior software engineer and expert in [programming language]. Provide a review of the following/attached code. Adhere to these requirements: 
    Format: [concise, bullet points, etc.]. 
    Tone: [educational, actionable, friendly, constructive, professional, etc.]. 
    Purpose: [identify all types of errors only, how to correct them, and explain why, identify errors only, identify errors and suggestions of improvements, etc.]. 
    Cognitive load: [requires minimal mental effort to process, comments should be easy to understand and actionable, or not relevant.]}
    
\end{center}

\medskip

Table~\ref{tbl:phase2_links} documents Phase II participants, the links to their LLM review, using the new prompt, and quotes from the followup interviews to illustrate their feedback on the new LLM's review.

\begin{table*}[t!]
\footnotesize

\caption{Link to ChatGPT 4o generated code review for Phase II followup interviews.}

\label{tbl:phase2_links}
\renewcommand\arraystretch{1.1}
\setlength{\tabcolsep}{6pt}

\begin{tabularx}{\textwidth}{l p{1.5cm} X}

\toprule

\textbf{\#} & \textbf{LLM review} & \textbf{Participant feedback on the LLM content}\\
\midrule

P1  & \href{https://chatgpt.com/share/68b1aa7c-da00-800c-bbe5-0530bfe24499}{Link}& \emph{``... prefers the new format [compared to Phase I]''} \\
P2  & \href{https://chatgpt.com/share/68b1724d-94ac-800c-9a74-682bab536060}{Link} & \emph{``This one is a lot better than the previous one. It's a lot more concise and really easy to read in the way that it's like sectioned. It is really nice as well.''}\\
P3  & \href{https://chatgpt.com/share/69130d16-ef1c-800c-b513-2a2b94a00d2f}{Link} & \emph{``This format is much better, easy to process.''}\\
P4  & \href{https://chatgpt.com/share/68b9ae50-1a2c-800c-ac82-19f1063d4be2}{Link} & \emph{``This one is better because I think it contains like the key points of the last one, but it's more compact, and to me, I find it easier to read.''}\\
P6  & \href{https://chatgpt.com/share/68b1e569-501c-800c-9c70-0c31b844e8ad}{Link} & \emph{``... this [ChatGPT 4o review] is very impressive and interesting.''}\\
P7  & \href{https://chatgpt.com/share/68b718bf-5654-800c-b966-7fd2cfe6b364}{Link} & \emph{``I like it, because it's very straight to the point.''}\\
P8  & \href{https://chatgpt.com/share/69130daf-a1bc-800c-8fac-eb2911fcd035}{Link} & \emph{``... they [errors] are valid things ... the older one [Phase I review] had more things.''}\\
P10 & \href{https://chatgpt.com/share/69130eb7-bc60-800c-b0ba-2b3ce1238c40}{Link} & \emph{``I was expecting these errors, spot on! And straightforward.''}\\
P12 & \href{https://chatgpt.com/share/69130f13-6598-800c-b23c-c3d231d4f2b4}{Link} & \emph{``Clean but I would have liked more details. But I can follow up with prompts.''}\\
P13 & \href{https://chatgpt.com/share/68b95379-f10c-800c-991c-ac8d673c33ce}{Link} & \emph{``I can see the verbosity is reduced with this prompt. I prefer this one, easy to digest.''}\\
P14 & \href{https://chatgpt.com/share/68b2c2cb-bef8-800c-8f70-572c6420f4fd}{Link} & \emph{``I think it's much better than last time ... It explains the problem concisely, then tells what the fix should be and displays an example. And does this a lot, and this is very helpful.''}\\
P16 & \href{https://chatgpt.com/share/69130fae-b630-800c-bbb3-18b693faa79e}{Link} & \emph{``Every section is well formatted ... I do not like the emojis! The best practices section is good but too short; we should have used educational.''}\\
P17 & \href{https://chatgpt.com/share/68b1c52c-2088-800c-8e40-aa4925135077}{Link} & \emph{``It matches pretty well. Honestly, yeah, it's pretty good.''}\\
P19 & \href{https://chatgpt.com/share/68b18b06-b650-800c-945b-c376833ad5be}{Link} & \emph{``It's shorter and easier to read.''}\\
P20 & \href{https://chatgpt.com/share/68b1b747-e79c-800c-b322-2afc22ae77b7}{Link} & \emph{``Right to the point, right? It's just the critical ones, and the second one, just for best practices. I would love it if somebody gave me this review.''}\\

\bottomrule

\end{tabularx}

\vspace{2mm}

\emph{Note:} Rows include only participants marked with an asterisk (*) in Table~\ref{tbl:population}, showing only Phase II follow-up participation.

\end{table*}

\subsection{Phase I \& II: Data Analysis and Integration}
\label{sec:analysis}

\begin{table*}[ht!]
\footnotesize
    \caption{Example of pattern codes and their corresponding \emph{First Cycle} codes}

    \label{tbl:themes}
    \renewcommand\arraystretch{1.8}
    
    \begin{tabular}{lp{2.8cm}p{7cm}}
    \toprule
      \textbf{Pattern codes} & \textbf{First cycle codes} & \textbf{Examples from the data}\\
      \hline
      
      \multirow{4}{*}{\shortstack[l]{\textbf{Self-regulation}\\ \textbf{strategies}}} & Reframing feedback  & \emph{``I try to not have any bad feelings about it, because in the end, I feel it is something that I can draw good things from and the aura of use from me. So I always try to, you know, be open towards criticism and not have, like, a negative emotional reaction to it''} (P4, Phase I).\\
      
     & Avoidance strategies & \emph{``So personally, it doesn't work with me, like, if someone is more like firm and maybe aggressive, I don't like that when people do that to me, so I try not to deal with other people''} (P16, Phase II).\\
      
      \hline
      
      \multirow{3}{*}{\textbf{Social calibration}} & Reciprocity norms& \emph{``... you know they're making an effort. They want to see you do well, and because they want to see you do well, you're inspired to do well.''} (P7, Phase I).\\
      
      & Relational climate & \emph{``I do not have a bad environment overall with my co-workers, so it's overall friendship wise and overall good and healthy. But that there could be some conflict in real life. And then you get actual comments, which are a bit harder in the language; depending on how healthy the relationship is''} (P1, Phase II).\\
      
     \bottomrule
     
    \end{tabular}
  
\end{table*}

\noindent We adopted a constructivist stance~\cite{creswell2016qualitative}, which aligns with our objective to understand software engineers' perceptions and  experiences with LLM-generated feedback compared to their peers. By adopting this stance, we recognize that knowledge and meaning are actively constructed by individuals through their experiences and interactions~\cite{creswell2016qualitative}.

We began by re-analyzing the Phase I interviews inductively. After completing the follow-up interviews, we analyzed the new dataset using a hybrid analysis approach: explicitly integrating themes derived from Phase I (deductive) with those emerging from data (inductive) ~\cite{fereday2006demonstrating}.

In Phase II analysis, we treated our existing Phase I pattern codes as a provisional coding framework~\cite{saldana2021coding,fereday2006demonstrating}. We applied this framework deductively to Phase II data while remaining open to inductively identifying codes for any new concepts emerging in the follow-ups. Each Phase I theme was used as a provisional code~\cite{saldana2021coding}; excerpts from Phase II were examined to determine their  degree of fit, expansion, or divergence. This allowed us to confirm and track the stability, expansion, or contraction of Phase I constructs over time. 
Inductively identified codes that did not map onto an existing Phase I pattern code were consolidated into new pattern codes.

We did not identify any divergence from Phase I themes; however, we enriched them with additional data. For \textbf{RQ4}, we identified new themes (e.g., aligned feedback and associated sub-themes). We also confirmed similar sense-making patterns for peer-led and LLM-assisted reviews; see Sect.~\ref{sec:findings} for detailed findings.

In both phases, we followed Miles et al.~\cite{miles2014qualitative} recommendations for the analysis process. First, we conducted a preliminary \emph{First Cycle} analysis. In this early stage of the analysis, we selected ``chunks'' of data that are pertinent to our RQs and assigned them codes~\cite{miles2014qualitative,saldana2021coding}. Using an inductive strategy for Phase I dataset and a hybrid for Phase II, this condensing yet analytical exercise allowed us to reduce the data corpus into meaningful concepts~\cite{miles1984qualitative}.

In the \emph{Second Cycle} that followed, we synthesized all the \emph{First Cycle} codes into ``Pattern codes''~\cite{miles2014qualitative}. In this integrative activity, we looked for patterns across the previous cycle's codes by identifying recurring themes that linked different codes together; whether through similarity, processual relationship, or complementary contributions to a cohesive construct~\cite{miles2014qualitative}. This process allowed us to develop higher-level constructs and abstract our understanding of the underlying phenomena. 

While the \emph{First Cycle} and \emph{Second Cycle} allowed us to generate constructs to answer our RQs, we needed to further understand the relationship amongst them. To this end, we used a ``causal network'' analysis~\cite{miles2014qualitative} to identify connections between the ``Pattern codes''. In this activity, we examined how ``Pattern codes'' influenced, conditioned, or intersected with one another in the data~\cite{miles2014qualitative}. These relationships allowed us to design our proposed models, i.e., figures~\ref{fig:rq1},~\ref{fig:rq2}, and~\ref{fig:rq3}, in Sect.~\ref{sec:findings}.

Table~\ref{tbl:themes} documents an example of some of our Pattern Codes and their corresponding \emph{First Cycle} codes. The final column provides examples from the data to illustrate the codes.

\subsection{Phase I \& II: Trustworthiness Techniques}
\label{sec:checking}

Reliability was ensured through iterative peer review of codes, consensus-building, and member checking~\cite{birt2016member}. Phase I and II coding cycles were conducted by the first author. Then, iteratively, the second and fifth authors reviewed the codes and Pattern Codes, provided comments, and proposed new and alternative codes, until a consensus was reached. The first author then revised and offered a final set of codes and Pattern Codes. This ``reliability check''~\cite{miles2014qualitative,creswell2016qualitative}, allowed us to validate our coding judgments and settle our discrepancies, resulting in more trustworthy interpretations.

During the re-analysis, we monitored thematic \emph{saturation}~\cite{morse2004theoretical,aldiabat2018data,miles1984qualitative,hennink2022sample}. We iteratively compared data and emerging themes to assess saturation points~\cite{morse2004theoretical}. Throughout the analysis, we managed to observe when our Pattern Codes reoccur strongly in the data~\cite{bowen2008naturalistic}, hence reaching saturation. We documented this process in our shared documents package. Miles et al.~\cite{miles2014qualitative} recommend thematic saturation to be evaluated at the \emph{Second Cycle} level, i.e., Pattern Codes or themes. We considered saturation reached when additional data did not yield new Pattern Codes or substantive extensions to existing ones~\cite{miles2014qualitative,saldana2021coding}. We share our saturation monitoring spreadsheet in our shared documents and data archive, see Sect.~\ref{sec:replication}. Most of our themes reached saturation at the 9th interview (Phase I) for \textbf{RQ2--RQ3} and follow-up interview (Phase II) for \textbf{RQ4}. \textbf{RQ1}'s saturation was not monitored directly, as it represents a high-level synthesis of the phenomena explored in \textbf{RQ2--RQ4}; its saturation is inherently evidenced through the underlying Pattern Codes of \textbf{RQ2--RQ4}.

Upon the completion of the analysis, we organized a \emph{member checking}~\cite{birt2016member} activity to collect feedback from our interviewees on our findings (see Sect.~\ref{sec:trust} for further details).

\paragraph*{Informed consent} Informed consents from the interviewees were obtained prior to the data collection in accordance with best practices and institutional requirements of the authors' institutions. We obtained consent in Phase I and II, including permission to record and transcribe the interviews, and to use anonymized quotations for research dissemination. Participants were informed about the purpose of the study, the procedures involved, the voluntary nature of participation, their right to withdraw at any time without consequence, and the measures taken to ensure confidentiality and secure data handling. Any identifying information in the transcripts was anonymized for reporting in publications. In cases where adequate anonymization was difficult or not possible, the corresponding data were excluded. Participants also agreed to publicly share the anonymized version of the interview transcripts and submitted code and reviews, as part of the  study's supplementary material. As part of the consent, participants also agreed that  their code would be used as input to prompt an LLM to generate reviews for it.

\subsection{Shared documents and data}
\label{sec:replication}

We share our data and other artifacts \href{https://doi.org/10.5281/zenodo.17597021}{here}\footnote{\url{https://doi.org/10.5281/zenodo.17597021}}. Interviewees consented to sharing anonymized interview transcripts. In this archive, we share:

\begin{itemize}

  \item [-]  \textbf{Phase I - Participants' code}: Source code submitted by participants.
  
  \item [-]  \textbf{Phase I - Peer-led code reviews}: Reviews authored by participants.
  
  \item [-]  \textbf{Phase I - ChatGPT 4o reviews}: LLM-generated reviews used in Phase I. For Phase II, the links to ChatGPT 4o reviews are available in Table~\ref{tbl:phase2_links}.
  
  \item [-] \textbf{Phase I \& II - Interview guides}: Protocols and question used for both phases in separate interview guides.
  
  \item [-] \textbf{Phase II - Prompt engineering experiment}: Report documenting the prompt evaluation procedure used in Phase II, including links to our tests.
  
  \item [-] \textbf{Data saturation}: Spreadsheet tracking saturation assessment across interviews.
  
  \item [-] \textbf{Member checking}: Questionnaire used for validation and the anonymized responses collected from participants.
  
  \item [-] \textbf{Phase I - Interviews}: Anonymized transcripts (all 20 participants).
  
  \item [-] \textbf{Phase II - Follow-up interviews}: Anonymized transcripts (excluding P3, P10, P12, and P14 due to anonymization risks).
  
  \item [-] \textbf{Codebook (combined Phase I \& II analysis)}: Pattern codes and \emph{First Cycle} codes, organized by \textbf{RQ2--RQ4}. (RQ1 omitted because it synthesizes RQ2--RQ4.)
  
\end{itemize}
\section{Findings}
\label{sec:findings}

\begin{figure*}[t!]

\centering

\includegraphics*[trim=0cm 0.60cm 0cm 2.6cm, clip, width=1.0\textwidth]{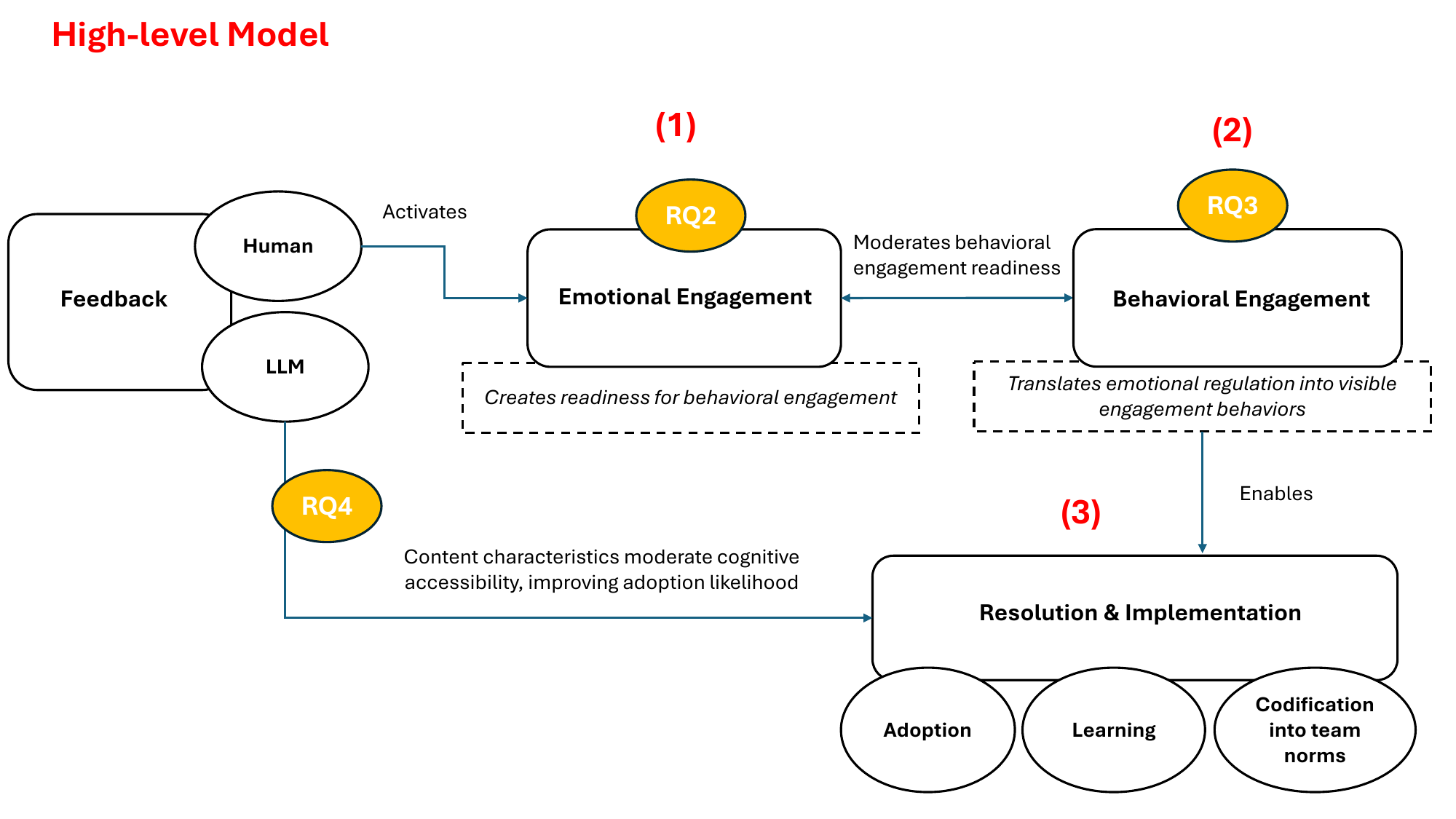}

\caption{High-level model (\textbf{RQ1}) of code-review engagement: emotional engagement $\leftrightarrow$ behavioral engagement $\rightarrow$ resolution and implementation (adoption, learning and team-norm codification); LLM content features moderate cognitive accessibility, improving adoption likelihood.}

\label{fig:rq1}

\end{figure*}

Recall our \textbf{RQ1}, which sought to identify the high-level engagement patterns of software engineers in human-led and LLM-assisted code reviews. Figure~\ref{fig:rq1} synthesizes our proposed high-level model of engagement in the context of code review. While our previous work~\cite{alami2025human} conceptualized the core dimensions of engagement and contrasted peer- and LLM-assisted reviews, in this study we unpacked the nuances of engagement. Our findings unfold a model shaped by three interdependent layers: \textbf{\emph{Emotional Self-regulation (RQ2)}}, \textbf{\emph{Behavioral Engagement (RQ3)}}, and \textbf{\emph{LLM Content Characteristics (RQ4)}}. These layers interact dynamically to influence the outcome of the review, i.e., \emph{Resolution \& Implementation}.

Our findings cohere around a dominant loop: \textbf{\emph{(1) Emotional Engagement}} $\rightarrow$ \textbf{\emph{(2) Behavioral Engagement}} $\rightarrow$ \textbf{\emph{(3) Resolution \& Implementation}}. This loop captures the prevailing sequence we observed; in practice, emotional self-regulation and behavioral engagement can be recursive and partially overlapping, especially when engineers regulate through dialogue (see Sect.~\ref{sec:rq2}--\ref{sec:rq3}). Stages (1)-(2) are more noticeable in human-led review, whereas (3) applies to both human-led and LLM-assisted contexts.

In the case of feedback from peers, when feedback is perceived as negative, engineers first self-regulate their emotions in response to feedback, which in turn moderates their readiness to engage; that readiness initiates a resolution stage (e.g., dialogue, negotiation) through cycles that may converge on the adoption of feedback's suggestions, often partially and selectively adopted after negotiation with the reviewer. As part of the resolution process, engineers may learn from the feedback, and in some cases, they may codify certain elements of it into shared team practices.

When engaging with review comments generated by an LLM, the interpersonal affect and social dynamics are reduced. However, the content characteristics of LLM reviews (structure, concision, tone, and ``why'' rationales) moderate cognitive accessibility of the review content and thus the likelihood of adoption. Despite the content styles being aligned with the engineers' preferences, the tension between the LLM's depersonalized neutrality, low emotional cost, and peers' interactional and relational needs persisted. The need for human connection, social validation~\cite{alami2024role}, and perceived human superiority in reviewing code also constrained some engineers' willingness to fully adopt LLMs as reviewers.

Using this high-level model, we align the subsections that follow with \textbf{RQ2 - Emotional Engagement}, \textbf{RQ3 - Behavioral Engagement}, and \textbf{RQ4 - LLM's Content Characteristics}, while distinguishing human-led review dynamics from LLM-assisted ones. Given that Fig.~\ref{fig:rq1} is an integrative level of our findings, we do not present separate evidence for \textbf{RQ1} in this subsection. Instead, \textbf{RQ1} is substantiated by the empirical results for \textbf{RQ2}--\textbf{RQ4}: emotional engagement (RQ2) and behavioral engagement (RQ3). In \textbf{RQ4}, we present how LLM content characteristics moderate cognitive accessibility and the likelihood of adoption. In the subsections that follow, we evidence each layer. When we use quotes from the interviews, we use Phase I or II in reference to which phase of the study the data was collected.

\subsection{RQ2 - Emotional Engagement}
\label{sec:rq2}

\begin{figure*}[t!]
\centering

\includegraphics*[trim=0.5cm 1.2cm 0.3cm 0.2cm, clip, width=1.0\textwidth]{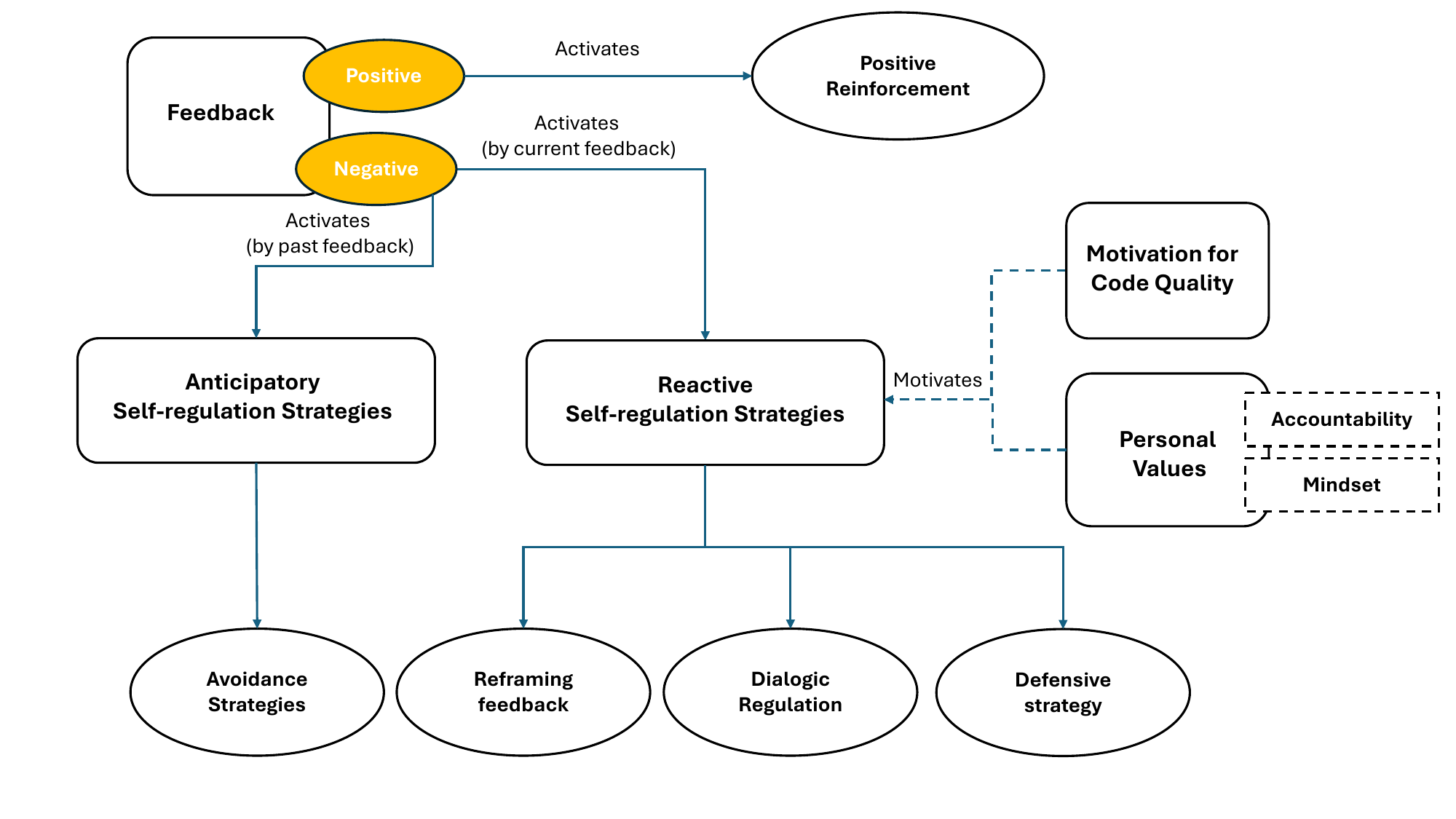}

\caption{\textbf{RQ2} - Emotional Engagement.}

\caption*{\footnotesize
\textbf{Legend:} Rounded rectangles denote constructs or strategies; ellipses denote transient states/phases. Solid arrows indicate activation flow; dashed arrows indicate value-driven influence. Amber ovals distinguish between feedback when perceived positive vs. negative.
}

\label{fig:rq2}

\end{figure*}

In response to \textbf{RQ2}, we identified a model for emotional engagement (i.e., Fig.~\ref{fig:rq2}). Our findings show that feedback from peers triggers an affective response that can elicit either negative (e.g., frustration, disappointment) or positive (e.g., pride, ``feeling good'') responses. When feedback is perceived as negative, engineers (code authors) seem to manage their emotions through a self-regulation process, the intrapersonal process through which they control their emotional responses. It is the hinge between feeling and doing; it appears to shape the engineers' behavioral engagement (\textbf{RQ3}) in the pursuit of a resolution for the feedback. 

In this regulation process, engineers draw on a repertoire of self-regulation strategies (e.g., reframing the feedback, engaging in dialogue, avoiding the reviewer, or, at times, becoming defensive) to resolve it. Some of these strategies are \textbf{\emph{Anticipatory}} (i.e., avoidance), enacted proactively to limit exposure, in anticipation of a known reviewer based on past experiences of perceived high-cost emotional encounters (e.g., harsh or overly picky reviews). On the other hand, reactive strategies (reframing, dialogic, and defensive) are activated upon receiving and reading the feedback.

Self-regulation strategies are value-driven (e.g., peer harmony, accountability) or anchored in engineers' motivation to improve their code quality. The outcome of this regulation phase is behavioral readiness, a state that may lead to a resolution. In sum, the findings of \textbf{RQ2} map a typical pathway from affective activation $\rightarrow$ self-regulation strategy (when feedback is perceived as negative) $\rightarrow$ positive reinforcement (when feedback is perceived as positive) $\rightarrow$ (in RQ3) behavioral engagement $\rightarrow$ resolution and implementation. However, as we discuss in Sect.~\ref{sec:rq3}, this transition is not strictly linear; when dialogic regulation is activated, self-regulation and behavioral engagement can unfold concurrently and in short cycles.

Depending on its formulation, when engineers perceive feedback as positive, it triggers a positive affective response, i.e., \textbf{positive reinforcement}; however, if it is deemed negative, then it activates a complex and often internal process of \textbf{emotional self-regulation}.

\paragraph*{Positive Reinforcement.} For authors, when framed positively and/or constructively, feedback seems to reduce self-regulation effort and may increase willingness to engage and accelerate the adoption process. For example, P10 appreciated that a reviewer started the review with a summary of the ``good'' aspects of his code: \emph{``... I see that at least it's telling me like good things, the good things about the code at the start, and then it's telling me all the things that could be improved. I really like that''} (Phase I, P10). Similarly, P17 echoed this sentiment regarding the positive feedback he received from P5: \emph{``makes me feel good. That always feels good''} (Phase I, P17).

For some engineers in our sample, positive reinforcement served as an interpersonal strategy to increase peers' receptivity to their feedback. P19 explains his approach as a reviewer: \emph{``... whenever I write a comment, and I try to not frame it negatively, like, instead of saying, don't do this, I would rather write something like I would write this and this instead, because whatever reason I have, ... whenever you approach someone with something negative, they will usually try to defend themselves instead of listening to what you're actually saying. So it's always a good thing to basically, come up with something that sounds more positive, that actually invites the other person to have a discussion ... making them more open to actually changing something''} (Phase I, P19). Positive framing of feedback is not only a strategy to influence code changes and peers' coding practices, it is also a social validation~\cite{alami2024role} mechanism and motivational tool. For example, when P5 was asked to explain his positive framing of comments, he replied: \emph{``I think positive feedback is important. I think both positive and negative feedback and finding a balance between the two. I think they're both important to keep people motivated and morale up and just you know, so people can take pride in their work and know that they did a good job''} (Phase I, P5). P12 puts it succinctly: \emph{``... if you can say, I thought it was very good that you wrote very short functions here, for example, then that person is more likely to keep writing short functions rather than sort of letting them ramble on .. [it] should help guide people to good behaviors, as well as steering them away from bad behaviors with sort of constructive criticisms''} (Phase I, P12).

In Phase I and II, engineers often described LLM reviews as consistently polite, and predictable. This consistency functioned as a form of positive reinforcement with low emotional cost, making comments easier to process and act on. These excerpts illustrate how this consistent tone shaped P4's and P11's engagement: \emph{``I like that it [the LLM] gives some positives. So it kind of reinforces my ideas in some way, like it, the structure is very good that it has like titles and then just points under each title, and makes it super easy to follow. Like I can quickly gain an overview, and I feel like I can quickly put this into practice ... I think it's quality feedback, honestly''} (Phase II, P4). \emph{``It was a nice thing to read [LLM's feedback], to say that, okay, I was correct. I did something correct. That's good, even though it's like feedback or changing stuff wasn't so harsh, like, I don't think it ever will be harsh with ChatGPT and, yeah, it just made it easier to kind of accept what it was telling me''} (Phase I, P11).

\paragraph*{Self-regulation strategies.} Emotional self-regulation is activated when feedback is difficult to receive or perceived negatively, e.g., \emph{``harsh''} (Phase I, P7 \& P10), \emph{``brutal''} (Phase I, P6), and \emph{``horrible''} (Phase I, P2). At this stage of engagement, engineers regulate their affect and felt quality of experience, which moderates their behavioral engagement during the resolution. 

Our data show a repertoire of strategies used by the engineers in our sample to self-regulate their emotions, spanning from \textbf{\emph{reframing the feedback}}, \textbf{\emph{dialogic regulation}}, \textbf{\emph{avoidance}}, and \textbf{\emph{defensiveness}}. These strategies seem to help engineers in our sample to manage the emotional affect and prepare to engage for resolution, enabling the shift from \emph{feeling} to \emph{doing} (i.e., \textbf{RQ3}). This does imply that \emph{feeling} categorically stops during the \emph{doing}. Emotions remain active and are continually reappraised when engineers negotiate meaning and implement changes; regulation may become concurrent with action. The self-regulation depicted in \textbf{RQ2}'s model is temporal and in response to reading and the subsequent perception of the feedback. Similar short and cyclic self-regulation episodes can recur as engineers advance the resolution and invest further in outcomes.

\textbf{Reframing.} The strategy is based on reinterpreting the intent of the feedback to reduce potential negative affect. As part of this reframing, our data show that engineers use task-anchoring and decouple themselves from the features of the comment (tone, style, phrasing, reviewer identity). They seem to dissociate the work that needs to be done from how the message was delivered, anchoring their attention in the errors and fixes instead.

When P10 and P12 were asked how they reacted to negative feedback they received from other participants, they described their reframing strategy: first, intent reinterpretation (the comments are to help rather than an attack); second, task-anchoring (centering attention on errors and improvements). In their words: \emph{``I don't take it personally or try not to take it personally ... It's just true [the errors are true] ... You always want to improve your work, don't you?''} (Phase I, P10); \emph{``... so I am aware of some of my triggers in that regard. I tried to spot when it happens and think, no, hang on. This isn't about me. This is just about a mistake that I made or a piece of knowledge that I didn't have. It's not an attack. For me, it's self-awareness of those triggers ... it's important to make sure that the code is correct before it gets released''} (Phase I, P12). P11 explained why he reframes: \emph{``it is very like when it starts to feel like a jab at you ... even if the person is coming at me with negative intent, I just try to make myself believe that they come positively. Otherwise I will get very hurt ... so I have to frame it in such a way, but it's easy for me to, you know, accept''} (Phase I, P11).

\textbf{Dialogic.} In this strategy, engineers use direct dialogue to reduce the tension caused by the negative feedback. They initiate a non-confrontational conversation (preferably 1:1, synchronous) to surface intent, clarify expectations, and re-establish common ground. Their belief is that human meaning is negotiated and not inferred from written comments. Thus, meaning is sought through direct exchange. P1 described this strategy as \emph{``you stay in the context''}; by bringing the conversation back to the shared situation, i.e., errors in the code, and not the tone and phrasing of the feedback. This strategy also uses task-anchoring. During this dialogue, when live artifacts are used (e.g., opened files, looking directly at the code), it may shift the focus on the situation and ``context'', lowering tension, and speeding the agreement on what to change. Notably, this strategy combines emotional self-regulation and resolution, while other strategies appear to be phased, i.e, self-regulation strategy $\rightarrow$ readiness to engage.

P1 explained the rationale and this approach succinctly: \emph{``... it's about a discussion, I need to know his idea what is, what the problem could be. I need his comments, and then we can overall, get a common ground, for example, okay, it makes no sense. I could interpret anything into it, and it could be wrong. That's why I'm definitely a guy which likes to discuss, to make it direct overall, because writing something in comments, I don't like it too much. It's more if you have a discussion, because it's faster. You stay in the context ... so there will be no misunderstanding, because in the meeting, you can look at faces and so on''} (Phase I, P1). In Phase II followup, he emphasized that prior to the decision to engage in a dialogue, he takes an internal decision: \emph{``you have to decide yourself if you want to, overall, have a conversation or just do it or ignore it. So you have to do some assessment yourself, and the decision''} (Phase II, P1).

In Phase I, P6 described the feedback he received from P11 as \emph{``brutal''}. Yet, he prefers to move from emotion-to-action using dialogue, seeking intent first (``what do you mean?'') before solution (``what should change?''): \emph{``I would say, for example, I would probably just ask them, what do you mean by that review. I would probably ask, do you mean that? So I would actually try to engage with them in a certain way, so that I try to understand exactly where they come from with their process as well. That's how we definitely have that discussion with them to understand exactly, okay, you went brutal. But at the same time, I'm gonna try to understand exactly what was the thought process in terms of how they approached it''} (Phase I, P6). His process rationalizes a ``brutal'' feedback into concrete tasks, by asking clarifying questions that shift the emphasis from the tone of the feedback to actionable changes.

Similarly, P7 perceived P11's review as \emph{``harsh''}. Yet, he is willing to engage in dialogue; when asked how he would deal with such a colleague, his strategy was dialogic: \emph{``... I'd still want to have a discussion with this reviewer. I would have my guard up a little bit more, because the way the tone comes across is a little bit more aggressive, so I kind of be a little bit wary, and I'd probably be careful with what I said too. I wouldn't want to further instigate things. Even if I disagreed, I would kind of definitely be careful choosing my words with this reviewer. Just to see, like, what the thought process was behind these comments, try to understand where this person is coming from''} (Phase I, P7).

\textbf{Defensive.} At times, feedback perceived as negative triggers defensive responses to reduce affect through self-protection and justification. For example: \emph{``it depends on the type of feedback. If I think it has merit, then sure I'll be willing to take it into consideration and make the change if I think it's appropriate. But if I have a disagreement with the feedback, I'll definitely stand up for it to make a case for why I think that way is better than the suggestions. Sometimes I can take it personally, like, if someone comes in with a suggestion that I don't agree with, it can be like personal''} (Phase I, P5).

P5's account shows a defensive self-regulation, where he manages emotional discomfort not by disengaging but by advocating for his decision and reaffirming personal competence. This strategy maintains engagement but may constrain openness to learning, functioning primarily as self-protection rather than a constructive one. While a dialogic strategy seeks meaning and clarification, a defensive one, as it appears in our analysis, moves directly to justification and advocacy, assuming the author's original solution is correct and the reviews' comments are a challenge to defend against rather than an opportunity to seek shared understanding. However, we do not claim that they could be mutually inclusive or employed consequentially, e.g., defensive $\rightarrow$ dialogic, as our data does not allow us to make such conclusion empirically.

\begin{table*}[t!]

\footnotesize
\caption{Self-regulation strategies: definitions, limitations, and impact on resolution. This analytic synthesis is intended as propositions and not as an exhaustively coded pattern or set of themes.}
\label{tbl:selfreg}
\renewcommand\arraystretch{1.25}

\begin{tabular}{p{3.1cm} p{4.2cm} p{5.5cm}}
\toprule
\textbf{Self-regulation strategy} & \textbf{Definition} & \textbf{Potential limitations \& impact on resolution}\\
\midrule

\textbf{Reframing feedback} &
Reinterprets the intent of the feedback (from negative to what needs doing), decouples tone, phrasing, and the reviewer's identity from the content, and anchors attention on the task (i.e., errors and fixes) to down-regulate defensive affect. &
May accelerate acceptance, reduce rumination, and speed the transition from \emph{feeling} to \emph{doing}. However, ambiguous or disputed items may remain unresolved or still require follow-up dialogue, a secondary strategy (i.e., dialogic).\\

\addlinespace

\textbf{Dialogic regulation} &
Seeks meaning in the feedback through 1:1 synchronous conversation to surface intent, clarify expectations, and re-establish common ground. &
Assumes a psychologically safe dialogue and carries the risk of encountering behavior similar to that observed in the written feedback, including potential escalation or conflict. The risk of delaying resolution is high.\\

\addlinespace

\textbf{Avoidance} &
Proactively limits exposure to emotionally high-cost reviewers. Participants reported deliberately not inviting ``toxic'' reviewers to review their code. &
May provide short-term relief; however, it assumes consistent effectiveness (i.e., that the engineer can reliably avoid the reviewers in question). Contingency strategies (e.g., dialogic or reframing) are needed when avoidance is not possible. It may lead to rapid resolution when successful.\\

\addlinespace

\textbf{Defensive} &
Manages emotions through control and advocacy rather than perspective-taking. &
May accelerate resolution when the dialogue is psychologically safe; however, it carries a high risk of escalation if the tone is harsh. It may also be perceived as combative, increasing interpersonal strain and reviewer ``pushback.''\\

\bottomrule
\end{tabular}
\end{table*}

\textbf{Avoidance.} This strategy is anticipatory, in contrast to the other self-regulation strategies, which are post-feedback. Engineers seem to proactively limit exposure to high-cost emotional sources (e.g., ``picky'', ``harsh'' reviewers) to protect affect. It is a premeditated (who to invite to a code review) rather than a post-feedback strategy like reframing. In the words of P6: \emph{``so in my environment, I try, by all means, to try to avoid adding that person to my reviews''} (Phase I, P6). P8 described some of his colleagues \emph{``picky''}. When asked how he deals with them, he explained: \emph{``in fact, there are only two of them and we know which one there are. They are always pushing, extremely picky. So we go for the ones that are not as picky''} (Phase I, P8).

The instrumental role of code review is to identify issues and actionable improvements and to converge on a resolution. The self-regulation strategies engineers use have distinct effects on resolution (and pace) and entail predictable limitations. Table~\ref{tbl:selfreg} synthesizes these impacts and constraints.

These strategies describe how engineers in our sample self-regulate their emotions in response to negative feedback. Our data also show that these strategies are value-laden or driven by the \textbf{\emph{motivation to achieve code quality}}. In our analysis, we identified three main values: \textbf{\emph{accountability}} to the team; a \textbf{\emph{mindset}} oriented toward growth and learning; and a commitment to \textbf{\emph{peer harmony}} (maintaining respectful, low-friction collaboration). These values inform and scaffold the strategies used (e.g., reframing, dialogue, avoidance) by the engineers. 

\textbf{Motivation to achieve code quality.} For many engineers in our sample, the motivation to achieve high-quality code functions as a driver to self-regulate in response to feedback. For some, quality is intrinsically driven, e.g., \emph{``I care''}, \emph{``so I try to be a person that cares because, well, it helps me from making mistakes. It is personal ... It's important to me that it [code quality] is done well''} (Phase I, P12). For P12, achieving code quality is entwined with her identity, e.g., \emph{``personal''}; so criticism is reframed as guidance toward a valued end (improved quality), rather than a threat. P13 motivation is also intrinsic but anchored in his personal integrity, e.g., \emph{``doing the right thing''}, and a desire for recognition for his standards. He explained: \emph{``just sort of that you're doing the right thing ... which makes me feel right ... having high-quality code that makes you feel like, Oh, I'm getting the high-quality code out there''} (Phase I, P13).

Others frame their motivation for quality in instrumental terms, \emph{``my main goal is to have a working program without bugs and with best practices''} (Phase I, P14). Collectively, these accounts show a value-goal system, which motivates self-regulating negative affect.

The same motivation appears source-agnostic and applies to LLM sources. Some engineers accept LLM input because it advances their pursuit of quality: \emph{``so as long the review is helping me to improve, I'm willing to consider a model like ChatGPT''} (Phase I, P8). This may imply that engineers evaluate LLM feedback through quality-achieving goals (i.e., does it identify errors? Propose best practices?), not a source lens, i.e., peers vs. LLM. Especially when the engineers perceive the emotional cost as lower when dealing with LLMs. P7 explained: \emph{``definitely this [ChatGPT's review]; I would choose this one just because it's organized, formatted, much better, and the tone is definitely more opening, more neutral, I guess ... I don't have to worry about if I would have a discussion with this person; I wouldn't worry as much about upsetting the other person ... I might be more picky with my words, just to make sure that it didn't get escalated into any kind of argument''} (Phase I, P7).

In short, motivation to achieve code quality operates as a driver of self-regulation in peer-led reviews. In LLM-assisted reviews, the same motivation calibrates engineers' openness towards the LLM and justifies adoption of suggestions.

\textbf{Accountability.} In peer-led review, accountability for code quality emerges at both the collective level (team) and the individual level. At the collective level, it seems a shared norm, a socially distributed sense of responsibility that binds engineers to one another and to the collective pursuit of code quality. When P6 was asked why they were invested in reviewing other participants' code, they explained: \emph{``so I think all of us are responsible. It's not only just the task ... So it is a collective thing that we all should be responsible for ... We also need to be accountable of what we do. So I like to be held accountable. Tell me when I'm wrong, tell me when I'm doing something correct''} (Phase I, P6). This shared accountability is both self-imposed, \emph{``because I'm accountable for my code and well, it's not only that, it's my work to do it''} (Phase I, P8).

In several accounts, this sense of responsibility appeared to orient self-regulatory responses (e.g., staying constructive, acting on issues) even when feedback delivery was imperfect. In this way, accountability functioned like a social contract in peer-led reviews. Engineers expected to be corrected and to correct, and framed action on feedback as ``part of the job''.

In the context of the LLM, accountability is individual and self-driven rather than collective. Engineers emphasize personal responsibility for verifying, contextualizing, and deciding whether to adopt LLM feedback. P9 explained: \emph{``I think I'm accountable dealing with the LLM. I always cross verify before just going and jumping ahead with what ChatGPT said''} (Phase I, P9). Similar attitude shown by P4: \emph{``so when I would be working with ChatGPT, then I would kind of also take on the responsibility that I basically have no one to blame but myself when things go wrong. But yeah, in general that is not the case in a team, we all accountable ... So yeah, again, if I decided to use ChatGPT for a task, then I am responsible for making sure that the output makes sense, and I just not plainly trust it''} (Phase I, P4). In this sense, accountability in LLM-assisted review reflects a shift from social obligation to moral responsibility; engineers remain accountable not to peers, but to their own professional standards.

\textbf{Mindset.} Across our sample, some engineers have shown a growth-oriented mindset in dealing with feedback. They approached feedback, regardless of whether it is negative, as an opportunity for learning, growth, and improvement rather than as a threat to their competence. When P20 was asked how they approach negative comment, they explained: \emph{``personally, I don't feel anything bad because I don't have that feeling that I'm always right, and I'm honestly open to new things, learning new things ... I think staying neutral is the key. Just basically do the job''} (Phase I, P20). For these engineers, feedback is a learning tool, not only for correctness but integrated prospectively, e.g., \emph{``because some of the things he said the next time in my next code, I will actually implement that, right? So, yeah, that's the thing, just to grow, improve yourself''} (Phase I. P3).

This mindset motivates engineers to stay open to criticism, approach the feedback as an opportunity to learn and translate it into concrete improvements, even when the tone or content are emotionally challenging.

In the case of the LLM-assisted review, this quality persists. Some engineers also approached the LLM review as a learning opportunity, e.g., \emph{``I definitely appreciate [LLM's review] being educational ... For example, it explained, how I handled this vulnerability well on the good aspects, and then explains what it was and how it will help prevent an attack. It is good. It reinforces good habits''} (Phase I, P10).

\begin{table*}[t!]
\footnotesize

\caption{Mapping of participants' self-regulation strategies and motivational supports. N in each column represent the number or occurrences of a particular strategy or motivational support in our data.}
\label{tbl:selfreg_map}

\renewcommand\arraystretch{0.90}

\begin{tabular}{l| c|  c|  c|  c | c| c| c }
\toprule

\multirow{2}{*}{\textbf{\#}} & 
\multicolumn{4}{c|}{\textbf{Self-regulation Strategies}} & 
\multicolumn{3}{c}{\textbf{Motivational Support}} \\

\cline{2-5}\cline{6-8}

& \textbf{Reframing} & \textbf{Dialogic} & \textbf{Avoidance} & \textbf{Defensive} & \textbf{Code Quality} & \textbf{Accountability} & \textbf{Mindset} \\

& \textbf{\emph{N=11}} & \textbf{\emph{N=13}} & \textbf{\emph{N=2}} & \textbf{\emph{N=2}} & \textbf{\emph{N=10}} & \textbf{\emph{N=14}} & \textbf{\emph{N=11}} \\
%\midrule
\hline
P1 &  & \textbf{\textbf{\checkmark}} &  &  &  &  \textbf{\textbf{\checkmark}} &  \textbf{\textbf{\checkmark}} \\
\hline

P2 & \textbf{\textbf{\textbf{\checkmark}}} &  &  &  &  & \textbf{\textbf{\textbf{\checkmark}}} & \textbf{\textbf{\textbf{\checkmark}}} \\
\hline

P3 &\textbf{\textbf{\checkmark}}  &  &  &  &  &\textbf{\textbf{\checkmark}}  & \textbf{\checkmark} \\
\hline

P4 & \textbf{\checkmark} &  &  &  &  & \textbf{\checkmark} & \textbf{\textbf{\checkmark}} \\
\hline

P5 &  &\textbf{\checkmark}  &  &\textbf{\checkmark}  &  &\textbf{\checkmark}  &  \\
\hline

P6 &  & \textbf{\checkmark} & \textbf{\checkmark}&  &  & \textbf{\checkmark} &  \\
\hline

P7 &  & \textbf{\checkmark} &  & \textbf{\checkmark} &  &\textbf{\checkmark}  &  \\
\hline

P8 &  & \textbf{\checkmark} & \textbf{\checkmark} &  &\textbf{\checkmark}  & \textbf{\checkmark} & \textbf{\checkmark} \\
\hline

P9 &  &\textbf{\checkmark}  &  &  &\textbf{\checkmark}  & \textbf{\checkmark} &  \\
\hline

P10 & \textbf{\checkmark} &  &  &  &\textbf{\checkmark}  &\textbf{\checkmark}  &\textbf{\checkmark} \\
\hline

P11 & \textbf{\checkmark} &  \textbf{\checkmark} &  &  & \textbf{\checkmark} &  & \textbf{\checkmark} \\
\hline

P12 & \textbf{\checkmark} & \textbf{\checkmark} &  &  & \textbf{\checkmark} &\textbf{\checkmark}  &  \\
\hline

P13 &  \textbf{\checkmark}&  &  &  & \textbf{\checkmark} &\textbf{\checkmark}  & \\
\hline

P14 & \textbf{\checkmark} &  \textbf{\checkmark}&  &  & \textbf{\checkmark} & \textbf{\checkmark} &\textbf{\checkmark}   \\
\hline

P15 &  &\textbf{\checkmark}  &  &  & \textbf{\checkmark} &  & \textbf{\checkmark}   \\
\hline

P16 & \textbf{\checkmark} &\textbf{\checkmark}  &  &  &  &\textbf{\checkmark}  &  \\
\hline

P17 &\textbf{\checkmark}  &  &  &  &  &  & \textbf{\checkmark}   \\
\hline

P18 &  &\textbf{\checkmark}  &  &  &  &  &\textbf{\checkmark}  \\
\hline

P19 &\textbf{\checkmark}  &\textbf{\checkmark}  &  &  &\textbf{\checkmark}  &  &\textbf{\checkmark}   \\
\hline

P20 &\textbf{\checkmark}  &  &  &  &\textbf{\checkmark}  &  & \textbf{\checkmark}  \\

\bottomrule
\end{tabular}
\end{table*}

Table~\ref{tbl:selfreg_map} synthesizes the distribution of identified self-regulation strategies across our sample. It reveals that engineers use multiple strategies, which may suggest a contingency approach, when the primary strategy is less efficient in a given context. Engineers may also combine strategies, e.g., reframing $+$ dialogic. Reframing and dialogic regulation emerge as the dominant and complementary strategies. They reflect two modes of self-regulation, intrapersonal (cognitive reappraisal and task-anchoring) and interpersonal (meaning negotiation, clarification, and task-anchoring). In contrast, avoidance appears as a selective, anticipatory mechanism aimed at minimizing emotional cost, while defensiveness surfaces rarely and primarily when feedback threatens personal competence. Both strategies combine dialogic either as a mitigation strategy, when avoidance is not possible, or combined with defensiveness. Across the table (Tbl.~\ref{tbl:selfreg_map}), these regulatory choices are motivated and scaffolded by a commitment to code quality, accountability, and a growth-mindset. Dialogic self-regulation is common where accountability is salient (e.g., P1, P6--P9, P12, P14--P16, P19). It seems that when code quality is a shared responsibility, engineers are willing to absorb the interpersonal cost of clarifying conversations to reach a resolution. The table also shows that growth-mindset orients from defensiveness and toward reframing and/or dialogic. In 13 of dialogic cases accountability is co-present, and in all of growth-mindset cases engineers adopt either reframing, dialogic or combined.

So, how does this intrapersonal process change when engineers process LLM-generated feedback? When the social-interpersonal layer drops out, engineers still draw on the same values, motivation for code quality, accountability, and growth mindset. But accountability becomes self-referential and not collective (team-level). In this setting, emotional self-regulation may become lighter and faster: reframing is rarely needed, avoidance becomes absolute as a strategy, and the main constraint to adoption is mostly cognitive (missing context, content style, and scope) and less emotional (see Sect.~\ref{sec:rq4}). As AI-supported collaboration grows, sustaining quality and ethical standards in software engineering will depend partly on engineers' ability to adapt these self-regulatory and value-based system to their work with machines. For example, while the values are transferable to an LLM-only setting, the self-regulation process must evolve, from managing social tension and affect toward calibrating trust, verifying reliability, and maintaining agency in decision-making.

\medskip

\medskip

\begin{mdframed}[linewidth=0.5pt,backgroundcolor=gray!06]

\noindent\textbf{RQ2 -- Summary (Emotional Engagement)}\\

\noindent\textbf{Trigger.} Peer feedback often functions as an affective trigger (positive or negative), prompting regulation, when perceived negative, before or during action. The trigger can be past exposure to reviewers (avoidance strategy) or the feedback received for a particular review (reframing, dialogic, defensive).\\

\noindent\textbf{Positive reinforcement.} When feedback is framed constructively or highlights positive aspects of the code, it evokes positive affect. Engineers interpret such feedback as acknowledgment of their competence, potentially sustaining their motivation to maintain similar performance and coding practices.\\

\noindent\textbf{Emotional self-regulation.} To manage negative feedback, engineers draw on four recurring self-regulation strategies: \emph{reframing} (intent reinterpretation $+$ task-anchoring), \emph{dialogic regulation} (meaning negotiation $+$ task-anchoring), \emph{avoidance} (anticipatory strategy), and sometimes \emph{defensiveness} (self-protection and advocacy for one's perspective in coding).\\

\noindent\textbf{Motivation and personal values as scaffolds.} Use of the above mentioned strategies is shaped by a \emph{motivation for code quality} and a value base: \emph{accountability} (team and individual), and \emph{growth mindset}.\\

\noindent\textbf{Function.} Self-regulation establishes \emph{readiness to engage} in the resolution process and channels movement from \emph{feeling} toward \emph{doing}.\\

\noindent\textbf{Patterns.} Reframing and dialogic self-regulation are most common, in our sample, and often complementary; avoidance is selective; defensiveness is rarer and less constructive.\\

\noindent\textbf{Bridge to resolution.} Readiness transitions into behavioral pathways (sense-making, resolution, implementation). However, the relation and transition are not always linear; they can be intertwined (\emph{self-regulation} $\leftrightarrow$ \emph{behavioral engagement + resolution}).\\

\noindent\textbf{LLM contrast.} In LLM contexts, emotional cost is significantly lower; values still guide engineers' actions, but accountability skews individual's decision-making for adoption pathways, explored in \textbf{RQ3} and \textbf{RQ4}.

\end{mdframed}

\subsection{RQ3 -- Behavioral Engagement \& Resolution}
\label{sec:rq3}

\textbf{RQ3's} findings trace how engineers move from \emph{feeling} to \emph{doing}, in response to the feedback. Engineers' emotional self-regulation seems to moderate their engagement readiness and may influence their behavioral pathway that follows: sense-making of comments, resolution, and then translation of decisions into concrete changes in the code. While emotional self-regulation often functions as preparatory phase to the process of resolution, it is largely intrapersonal to establish readiness. However, as demonstrated by Table~\ref{tbl:selfreg_map} summary, the relationship between emotional self-regulation and behavioral engagement is not always intrapersonal and linear, especially when dialogic strategy is used, either alone (e.g., P1 and P18), as contingency, or in combination with other strategies (e.g., P6, P8, and P12); the self-regulation and engagement become intertwined. In such cases, emotional self-regulation may unfold simultaneously with interaction. Engineers appear to self-regulate their affect through the dialogue itself, using conversation to clarify intent, restore mutual understanding, and convert tension into collaborative action.

When engineers choose to respond, whether part of the dialogue or otherwise, they seem to calibrate their behaviors according to the social context of their teams. They align their behavior to the relational climate with the reviewer or the team norms, how other peers would engage in such a situation, or what is acceptable by the team. This social calibration informs resolution behaviors (e.g., negotiation, consensus-seeking, decision-making). Then, the implementation of changes. In sum, \textbf{RQ3's} findings map the \emph{doing}: from social calibration (when responding to feedback) $\rightarrow$ resolution $\rightarrow$ implementation. However, in the case of LLM-assisted review, the decision process is largely unilateral, internal, and influenced partly by the content style (see \textbf{RQ4}, Sect.~\ref{sec:rq4}).

\textbf{Social calibration.} In this process, engineers adjust their response tone, style, phrasing, etc. to take social factors into account or to allow alignment and comparison with other peers. This calibration appears to be influenced by the \textbf{relational climate with the reviewer}, \textbf{team norms}, and/or \textbf{social comparison}.

\begin{figure*}[t!]

\includegraphics*[trim=0cm 2.5cm 0cm 1.5cm, clip, width=1.0\textwidth]{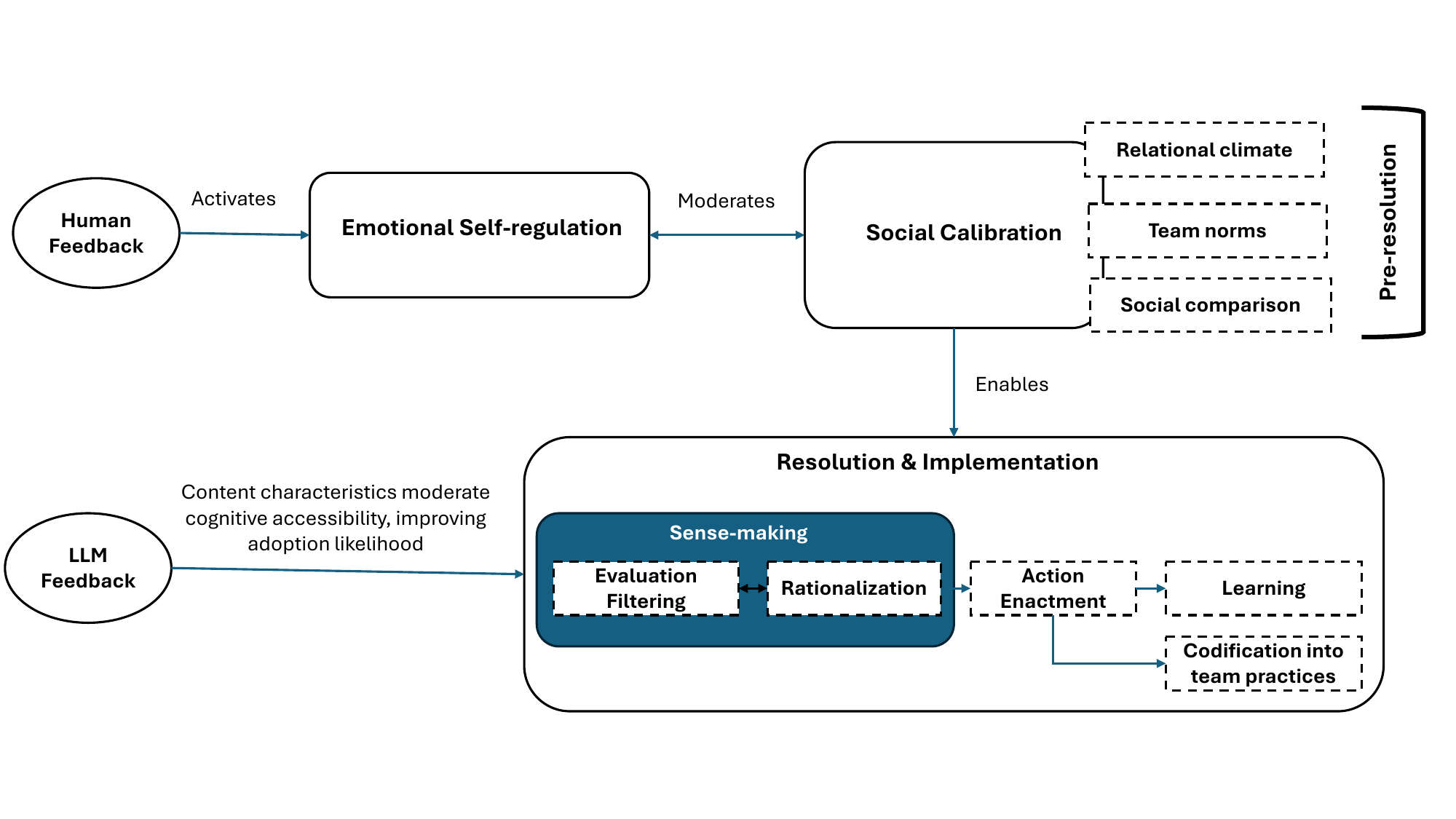}

\caption{\textbf{RQ3} - Behavioral Engagement and Resolution. Emotional Self-regulation is the model described in Sect.~\ref{sec:rq2} (RQ2).}

\caption*{\textbf{Legend:} Ovals distinguish between feedback from peers vs. LLM. Rounded rectangles denote constructs with components (e.g., \emph{Evaluation Filtering}, \emph{Rationalization}, \emph{Action Enactment}); dashed rectangles denote components. Solid arrows indicate activation flow. The blue rounded rectangle denotes the internal \emph{sense-making} process of the feedback used by software engineers part of the decision-making to adopt the comments.}

\label{fig:rq3}

\end{figure*}

\textbf{Relational climate.} The perceived quality of interpersonal relations moderates how engineers choose to engage in resolution. For example, when P16 was asked why he would choose to remain constructive with the reviewer, he explained: \emph{``I feel like you create like this toxic workplace vibe ... sometimes you create relationships with people like friends, not just co-workers. And then if you start being too tough on them, you kind of lose that''} (Phase I, P16). To engineers in our sample, relational climate is a proximal cue that shapes how they interpret and engage to reach resolution. P15 explained: \emph{``I prefer a friendly approach ... if the review comes from someone I know in real life, of course, this kind of interaction also helps to digest the feedback and improve your personal relationship, whether you are receiving a good or bad review to improve your code. I know that person cares about me and what I'm doing''} (Phase I, P15).

These accounts explain a drive for peer harmony as a mechanism to maintain relational climate. As P6 explained, \emph{``I think it's very important to build a good working relationship with people... Even though we might be laughing together when we're having lunch... we also need to be very constructive in terms of our work''} (Phase I, P6). Likewise, P10 emphasized sustaining harmony: \emph{``... also want to just keep a good relationship with peers ... even if they really don't like me, I'll still be respectful to them''} (Phase I, P10).

\textbf{Team norms.} Our data support two complementary team-norm mechanisms: (1) \textbf{psychological safety}, engineers can ask, push back, and be heard without interpersonal penalties, and (2) \textbf{reciprocity of effort}, reviewers' labor justifies constructive behavior and similar investment in effort.

\textbf{Psychological safety.} Our data show that engineers feel psychologically safe when they perceive having a voice without penalty. They can question, push back, and voice concerns without censure. P1 explained his approach to reduce avoidance dynamics (withholding PRs, delay, not asking for help) and keep review interactions safe to enable improvement: \emph{``I don't go the hard, harsh way. You have to be friendly ... they are more likely to make PR pull requests. If you get always harsh comments, you will not ask for help. That's why I also like to make it more friendly, more discussion. Hey, why did you do that? Let's discuss it''} (Phase I, P1). In his team, P11 explained that it is safe to push back: \emph{``push back is not discouraged''} (Phase I, P11). P7 explained that in his team, it is acceptable to make ``mistakes'', \emph{``let's say somebody made a mistake and I had to review it, or I made a mistake, somebody was telling me the feeling that okay, it's okay to make that mistake, and it's okay to fail, and we'll help you through it. We're all working together, try to make our application work and do what the business needs, just that feeling of working together, that cooperation, I think that's very key to like being a working team of developers''} (Phase II, P7).

\textbf{Reciprocity of effort.} The engineers in our sample also discussed a reciprocity norm: if their peers invest time to review, then they feel that they owe them constructive engagement. Engineers perceived invested effort in reviewing their code as an invitation to constructive response. P4 explained: \emph{``... people that write these criticisms also invest their time and effort into these reviews, and in the end, they help me. So I think it's only fair to kind of react to that with an open mind''} (Phase I, P4).

\textbf{Social comparison}. Our data also show that engineers calibrate their engagement through social comparison, evaluating their own performance against peers' contributions. For example, P2 described a comparison with P19: \emph{``... when I looked at that, I felt a bit ashamed of what I did for P19. So by, you know, the code that they review, that P19 did for me is a lot better than what I did for them. I was very impressed by what they put onto it''} (Phase I, P2). Then later in the interview he drew parallel with his current team: \emph{``It inspires me. Like, it really inspires me like, you know, I see that sometimes with some of the colleagues. They go above and beyond ... I should be more like that ... someone cares, you know they're making an effort. They want to see you do well, and because they want to see you do well, you're inspired to do well''} (Phase I, P2). P2's account shows adaptive striving rather than defensiveness. His peers set an internal benchmark for him for quality and professionalism.

\textbf{Resolution \& Implementation.} In our model (Fig.~\ref{fig:rq3}), emotional self-regulation and social calibration function as preparatory but continuous processes. They guide when and how engineers move toward the resolution stage. Emotional self-regulation seems to prepare the individual to engage by stabilizing affect, while social calibration may orient the response within the relational and normative landscape of the engineers' teams. 

Our data indicate that engineers will eventually transition to resolution, regardless of the outcomes of their intrapersonal self-regulation and interpersonal engagement, if any, such as dialogic interactions or responses to comments. Some engineers, in our sample, show an inclination to process the feedback unilaterally using an internal \emph{sense-making} process followed by a selective adoption of comments. However, none of the engineers indicated signs of potential withdrawal, the behavioral disengagement from the review process, by avoiding or ignoring feedback, delaying responses, or refraining from participation to protect oneself from emotional or interpersonal strain~\cite{carver2004self,gross1998emerging}. 

This observation in our sample may reflect the controlled nature of our study. Participation was structured and time-bounded, reducing opportunities for avoidance. Additionally, participants may have experienced limited psychological threat in the review context, since the feedback was not tied to real workplace experience (discussed further in Sect.~\ref{sec:trust}).

Resolution is not merely about accepting or rejecting feedback; it is a process of \textbf{sense-making}, and decision-making. Some engineers seek clarification or consensus (e.g., through dialogic exchanges), while others act independently once understanding is reached. The underlying aim is convergence, aligning perspectives on what needs to change and why. Then, implementation follows, i.e., edits to the code. This process closes the feedback loop that began with emotional self-regulation.

\textbf{Sense-making.} The sense-making process is either unilateral (e.g., framing strategy alone) or dialogic, when the engineer chooses a direct interaction. It comprises two intertwined and parallel processes, \emph{\textbf{evaluation filtering}} and \emph{\textbf{rationalization}}, before acting on the comments, i.e., \emph{\textbf{action enactment}}. P8 explained this process: \emph{``most of the time, if a comment is about an error or an edge case that doesn't work, I just fix it -- 100\%, no question. But for the rest of the suggestions, I first read and check if they make sense to me. If they do, I try to explain my reasoning, why I did it that way, and see if the other person agrees. Because most of the time, they write the comment without knowing my thought process. So I explain, I did this because of that, and if they have another view, I can change it, no problem. But if I think the comment isn't valid or doesn't apply, I do the same thing -- just with more arguments to explain why I wouldn't change it''} (Phase I, P8).

His process start with \emph{\textbf{evaluation filtering}}, \emph{``I check if they make sense to me''}, and rationalization, \emph{``I try to explain my reasoning''}, followed by \textbf{\emph{action enactment}}, \emph{``I fix it''}, or \emph{``I wouldn't change it''}. However, rationalization is not always dialogic. Some engineers engage in this process internally, weighing the rationale of the comment against their own internal understanding of the context of the code and coding practices. As P14 explained: \emph{``I would check the comments and the reviews and try to improve the code given the review ... For the ones I think they are correct, yes, for the other ones, I would have to think about it if they make sense in this case or not ... Or the context of the code I'd like to see if they apply or not? For example, I see here a comment because there are some methods, some getters, that are not used. I think I would keep them because they could be used elsewhere''} (Phase I, P14).

\textbf{Learning.} Code review is not only to correct errors and improve the code. Engineers in our sample leverage the process to learn. In the words of P11 and P8: \emph{``early in my career, I had a very good mentor who kind of taught me a lot, especially through code reviews. That's how I learned a lot of how to improve myself. And I always feel like I need to give back, you know, to also help others improve their code''} (Phase I, P11); \emph{``... new ideas ... I have learned a lot} (Phase I, P8).

\textbf{Codification.} Our data show that in some instances resolutions may require a team consensus. Engineers may opt to transform their code review resolutions into shared practices. Codification represents the social consolidation of learning; negotiated solutions, stylistic preferences, or best practices evolve into agreed-upon standards at the team-level. Behaviorally, it reflects a shift from individual cognition (``how I fix it'') to collective alignment (``how we do it''). 

In our data, this process is deliberative. Engineers describe codification as an outcome of discussion, argumentation, and consensus-making. P1 illustrated this collective reasoning: \emph{``it's not like, hey, you have to follow my path ... and if we can't find common ground, let's vote for it ... because the team has to, overall, maintain the code, for example, not only a single person''} (Phase I, P1). Once a resolution gains sufficient acceptance, it becomes embedded as a team practice or guideline, as P1 later explained, \emph{``so the team has to decide which way we have to go. ... if we can generalize it, make it more generic, then we can edit it as a guideline that we have to do a certain way''} (Phase II, P1). Similarly, P9 highlighted the integrative function of this process: \emph{``it's how the work becomes seamless if the team's all in one page ... It's good to have a discussion and come on to the same page''} (Phase I, P9).

\begin{landscape}

\begin{table}[t!]
\footnotesize

\caption{From self-regulation to closure: mapping strategies, motivation, and outcomes.}
\label{tbl:traject}

\renewcommand\arraystretch{1.0}

\begin{tabular}{l|c|c|c|c|c|c|c|c|c|c|c}
\toprule

\multirow{2}{*}{\textbf{\#}} &
\multicolumn{4}{c|}{\textbf{Self-regulation strategies}} &
\multicolumn{3}{c|}{\textbf{Motivational \& values support}} &
\multicolumn{1}{c|}{\textbf{Resolution path}} &
%\multicolumn{1}{c|}{\textbf{Locus of engagement}} &
\multicolumn{1}{c|}{\shortstack{\textbf{Locus of}\\\textbf{engagement}}} &
\multicolumn{1}{c|}{\textbf{Outcome}} &
\multicolumn{1}{c}{\textbf{Pace}} \\
\cline{2-8}

& \textbf{Reframing} & \textbf{Dialogic} & \textbf{Avoidance} & \textbf{Defensive} & \textbf{Quality} & \textbf{Acc.} & \textbf{Mindset} & \textit{process} & \textit{who} & \textit{implementation} & \textit{speed} \\
\midrule

P1  &  & \checkmark &  &  &  & \checkmark & \checkmark & Negotiate $\rightarrow$ Escalate & Dyad $\rightarrow$ Team& Partial & Moderate \\
\hline

P2  & \checkmark &  &  &  &  & \checkmark & \checkmark & Internal sense-making  & Solo & Partial & Fast \\
\hline

P3  & \checkmark &  &  &  &  & \checkmark & \checkmark & Internal sense-making & Solo & Partial & Fast \\
\hline

P4  & \checkmark &  &  &  &  & \checkmark & \checkmark & Internal sense-making  & Solo & Partial & Fast  \\
\hline

P5  &  & \checkmark &  &  \textbf{\checkmark} &  & \checkmark &  & Negotiate & Dyad & Partial & Slow \\
\hline

P6  &  & \checkmark & \checkmark &  &  & \checkmark &  & Negotiate $\rightarrow$ Escalate & Dyad $\rightarrow$ Team& Partial & Slow \\
\hline

P7 &  & \checkmark &  & \textbf{\checkmark} &  & \checkmark &  & Negotiate $\rightarrow$ Escalate  & Dyad $\rightarrow$ Team & Partial &Slow  \\
\hline

P8  &  & \checkmark & \checkmark &  & \checkmark & \checkmark & \checkmark & Negotiate & Dyad & Partial & Moderate \\
\hline

P9  &  & \checkmark &  &  & \checkmark & \checkmark &  & Negotiate & Dyad &Partial  &  Moderate\\
\hline

P10 & \checkmark &  &  &  & \checkmark & \checkmark & \checkmark & Internal sense-making & Solo & Partial & Fast \\
\hline

P11 & \checkmark & \checkmark &  &  & \checkmark &  & \checkmark & Negotiate & Solo $\rightarrow$ Dyad & Partial & Moderate \\
\hline

P12 & \checkmark & \checkmark &  &  & \checkmark & \checkmark &  & Negotiate & Solo $\rightarrow$ Team & Partial & Slow \\
\hline

P13 & \checkmark &  &  &  & \checkmark & \checkmark &  & Internal sense-making& Solo & Partial & Fast\\
\hline

P14 & \checkmark & \checkmark &  &  & \checkmark & \checkmark & \checkmark & Negotiate & Dyad & Partial & Moderate \\
\hline

P15 &  & \checkmark &  &  & \checkmark &  & \checkmark & Negotiate &Dyad  &Partial  &  Moderate\\
\hline

P16 & \checkmark & \checkmark &  &  &  & \checkmark &  & Negotiate & Dyad &Partial  & Moderate \\
\hline

P17 & \checkmark &  &  &  &  &  & \checkmark &  Internal sense-making& Solo & Partial & Fast\\
\hline

P18 &  & \checkmark &  &  &  &  & \checkmark & Negotiate &Dyad  & Partial & Moderate \\
\hline

P19 & \checkmark & \checkmark &  &  & \checkmark &  & \checkmark & Negotiate & Dyad & Partial & Moderate \\
\hline

P20 & \checkmark &  &  &  & \checkmark &  & \checkmark & Negotiate & Solo & Partial & Fast \\

\bottomrule

\end{tabular}

\vspace{6pt}
\raggedright
\footnotesize 

\textbf{Legend:} 

Self-regulation strategies include Reframing, Dialogic, Avoidance, and Defensive.\\

Motivational and values support categories include Code quality (i.e., Quality), Accountability (i.e., Acc.), and Mindset.\\

Resolution path: \textbf{Internal sense-making} - closure without negotiation and/or escalation; the feedback is evaluated through an individual sense-making process. / \textbf{Negotiate} - co-construction of meaning before closure/implementation, mostly used in a dialogic self-regulation. / \textbf{Escalate} - seeking broader consensus; in our data, it is escalating to team-level and conditional to the outcome of the dyad negotiation. / Defer / Ignore.\\

Locus of engagement: with whom engineers direct their engagement when responding to feedback; Solo / Dyad / Team or combined sequentially.\\

Outcome: Adoption (full adoption) / Partial / Rejected. Full and partial adoption are contingent to the negotiation (e.g. dialogic) or internal sense-making (e.g., reframing only).\\

Pace: Fast / Moderate / Slow.\\

\end{table}

\end{landscape}

To deepen our understanding of the patterns identified in the analysis, we synthesized each participant's trajectory from self-regulation to resolution and outcome in Table~\ref{tbl:traject}. For self-regulation strategies and motivational and/or values supports (e.g., code quality, accountability, or team norms), they were directly coded from participants' interviews. For pace and resolution outcome, they were inferred analytically, such as the degree of decisiveness, negotiation, or iteration, as described in the interviews. This mapping allows us to trace how emotional, social dynamics and subsequent behavioral engagement shape the resolution process and its eventual closure. For example pace was inferred as ``Moderate'' in the case of P1 because he prefers to negotiate and escalate if needed. For P20, it is ``Fast'' because they indicated preference for reframing only.

Table~\ref{tbl:traject} shows an alignment between the locus of engagement and the pace of resolution. When engineers resolve feedback through solo and internal sense-making (6 out of 20 cases), the process could be fast. In contrast, dyadic negotiation may lead to a moderate pace, with resolution bounded by the consensus of the pair. The slowest trajectories appear when feedback escalates to the team level (e.g., P1, P6, P7), yet this practice is necessary when shared standards and collective ownership are the norms. Notably, partial adoption is the standard, reflecting the selective nature of processing the feedback and its dependency on negotiation and engineers' internal sense-making.

However, fast may not necessarily imply thorough, and may cause a higher risk of idiosyncratic implementation and weaker alignment with peers' expectations. In addition, dialogue does not guarantee consensus. Dyads may stall, cause friction, and require escalation to team-level governance. In sum, the trajectory to resolution is fluid and contingent on situational contexts and social settings rather than a universal formula.

So, how do these trajectories to closure change when engineers interact with an LLM? Our data show that engineers may default to internal sense-making, with faster resolution, implying potential higher risk of idiosyncratic implementation and a weakening of team norms, including collective standards (see Sect.~\ref{sec:rq4}).

\medskip
\medskip

\begin{mdframed}[linewidth=0.5pt,backgroundcolor=gray!06]

\noindent\textbf{RQ3 -- Summary (Behavioral Engagement \& Resolution)}\\

\noindent\textbf{From \emph{feeling} to \emph{doing}.} Emotional self-regulation establishes \emph{readiness} and often precedes action; when dialogue is chosen, regulation and engagement intertwine (engineers regulate \emph{through} the interaction).\\

\noindent\textbf{Social calibration.} Before acting, engineers attune to: (a) \emph{relational climate} (maintain peer harmony, avoid toxicity), (b) \emph{team norms} (psychological safety; reciprocity of effort), and (c) \emph{social comparison} (adaptive striving toward peer standards), not necessarily used alone, but combining multiple.\\

\noindent\textbf{Resolution.} Resolution involves a sense-making process (either individual/internal or during the dialogue). It comprises \emph{evaluation filtering} and \emph{rationalization} of the comments (solo or dialogic), followed by \emph{action enactment} (edits of the code, replies to the comments, or both).\\

\noindent\textbf{Resolution paths.} Typical paths include \emph{internal sense-making} (solo closure), \emph{negotiate} (dyadic meaning-making), and \emph{escalate} (team consensus when standards and trade-offs are at stake). However, sometimes these are combined, e.g., P9 and P12 in Tbl.~\ref{tbl:traject}.\\

\noindent\textbf{Locus alignment with resolution pace.} Solo/internal paths maybe be \emph{fast} but idiosyncratic; dyadic negotiation could be \emph{moderate}; team escalation may delay resolution, i.e., \emph{slow}, yet supports shared standards.\\

\noindent\textbf{Outcomes.} \emph{Partial adoption} is common, reflecting item-by-item uptake and negotiated closure rather than collective acceptance of comments. It may also imply that code quality and coding decisions are negotiable rather than rigid standards.\\

\noindent\textbf{Learning \& codification.} Individual fixes can consolidate into individual learning and/or team guidelines when comments and improvements warrant a team-level discussion and agreement.\\

\noindent\textbf{LLM contrast.} In LLM-assisted review, decisions skew \emph{solo/internal}; emotional ease and social cost can shorten time to action, but risks drift from team norms unless practices are re-aligned (see Sect.~\ref{sec:rq4}).

\end{mdframed}

\subsection{RQ4 -- LLM's Content Characteristics}
\label{sec:rq4}

\begin{table}[th!]
\footnotesize

\setlength{\tabcolsep}{4pt}

\caption{Aligned LLM feedback: content features and cognitive effects.}
\label{tbl:rq4_aligned}

\renewcommand\arraystretch{1.0}

\begin{tabular}{p{2.3cm} p{2.2cm} p{8.2cm}}
\toprule

\textbf{Feedback feature} & \textbf{Cognitive effect} & \textbf{Illustrative evidence (Phase II)}\\
\midrule

\multirow{4}{*}{Structure} & \multirow{4}{*}{\emph{``Easy to read''}} & \emph{``Yeah, this one is a lot better than the than the previous one. It's a lot more concise and really easy to read in the way that it's like sectioned is really nice ... by doing that, you kind of reduce, like, the cognitive load on the human.''} (Phase II, P2).\\

\addlinespace

\multirow{2}{=}{Why-based justification} & \multirow{2}{*}{\emph{``Easy to read''}} & \emph{``... in a code review, you kind of just want to know, where's it going wrong? What am I missing? Am I missing tests? If it's been inefficient, why, and how can I make it more efficient?''} (Phase II, P2).\\

\addlinespace

\multirow{4}{*}{Actionability} & \multirow{4}{*}{\emph{``Easy to read''}} & \emph{``I'm kind of thinking about if what I'm reading makes sense. This one is better because I think it contains like the key points, but it's more compact. I find it easier to read ... I definitely intend to keep using it. I think it is useful.''} (Phase II, P4).\\

\addlinespace

\multirow{2}{*}{Tone} & \multirow{2}{*}{\emph{``Trust''}} & \emph{``I like the tone for sure, very neutral. It's not very positive and supportive. So it's very professional ... because it's very straight to the point, so I think that trust would be built up more.''} (Phase II, P7).\\

\addlinespace

\multirow{4}{=}{Fit-for-purpose scope} & \multirow{4}{*}{\emph{``Right to the point''}} & \emph{``I think it's done a pretty good job. So it found all of the critical errors that could fail if it was in production. And, yeah, it also gave the small, small fixes as well. So yeah, I would say, I like it ... Right to the point. I would love it if somebody gave me this review.''} (Phase II, P20).\\

\bottomrule
\end{tabular}
\end{table}

Recall \textbf{RQ4} sought to investigate how the content characteristics of LLM-generated feedback (e.g., scope, format, and tone) shape software engineers' engagement and willingness to adopt. Our Phase II prompt (see Sect.~\ref{sec:methods}) has improved the alignment with the engineers' preferences in our sample; e.g., \emph{``honestly like to get feedback like this from one of my coworkers. I would be pretty shocked''} (Phase II, P7); and \emph{``This is very impressive and interesting''} (Phase II, P6). Thus, the question becomes not how engineers regulate emotion and their subsequent behaviors pursuing resolution, but how they act on feedback. The emotional cost and the need for self-regulation are significantly reduced in LLM-assisted review; e.g., \emph{``... human beings can be very taxing emotionally ... I think AI kind of has a better element for me because of the fact that the emotion part is actually removed. I only get the response part, where you actually just need to know exactly how you can resolve the issue ... with AI, I get unbiased, unfiltered criticism''} (Phase II, P6).

Our analysis identified two outcomes: improved engagement and likelihood of adoption when the \textbf{feedback is aligned} with the engineers' preferences, and potential disengagement when the \textbf{feedback is unaligned}.

\textbf{Aligned feedback.} Aligned feedback is when LLM-generated feedback matches engineers' cognitive expectations and is scoped accordingly. When feedback aligns with format, actionability, tone, and fit-for-purpose preferences, engineers in our sample showed higher engagement and improved willingness to adopt. Table~\ref{tbl:rq4_aligned} illustrates the alignment and its impact on engagement. Aligned feedback was associated with lower processing effort and engineers often reported moving more quickly to implementation.

Across cases, engineers linked cognitive ease (clear structure, concise scope, neutral tone) to lower effort in sense-making and higher stated likelihood of adopting the suggestions; nonetheless, likelihood of adoption remained conditional on perceived correctness and context fit rather than ease alone. P19 explained how she would process the LLM's feedback: \emph{``I would look at it one by one and see whether I think that is useful or not, all the useful ones, I would implement them right away and everything else where I still have questions or I could provide more context. I would do that later''} (Phase II, P19). This account mirrors the \textbf{internal sense-making pathway} described in \textbf{RQ3}: engineers filter, rationalize, and then enact actions when cognitive effort is low and task-feedback alignment is high. In other words, LLM-generated feedback that minimizes cognitive friction seems to reduce the move from comprehension to implementation, at the same time reducing the perceived need for additional dialogue or emotional self-regulation.

\begin{table}[t!]
\footnotesize

\setlength{\tabcolsep}{4pt}

\caption{Unaligned LLM feedback: content features, cognitive friction, and disengagement patterns.}

\label{tbl:rq4_unaligned}
\renewcommand\arraystretch{1.05}

\begin{tabular}{p{3.2cm} p{2cm} p{2cm} p{5.2cm}}
\toprule

\textbf{LLM feedback feature} & \textbf{Cognitive effect} & \textbf{Disengagement} & \textbf{Illustrative evidence (Phase I)} \\

\midrule

\multirow{8}{=}{Verbosity / over-explaining} & \multirow{8}{=}{\emph{``I need to filter this out''}} & \multirow{8}{=}{\emph{``It's not useful''}} & \emph{``But not the verbose thing too is too much. It's most likely written for somebody else, because the other person could be not knowing what's happening overall ... for me, it's not useful because I know it. I wrote it overall. I know what it does, and it's too much. It's the information which is obsolete, in my opinion. So, of course, there is certain improvements available. This is very good, but it's too verbose. It's giving me too much information, which I already know. Then I need to filter this out''} (Phase I, P1).\\

\addlinespace

\multirow{6}{=}{Generic or template-like tone} & \multirow{6}{=}{\emph{``makes no sense to me''}} & \multirow{6}{=}{\emph{``it doesn't give me value''}} & \emph{``Yeah, what I don't like overall is that it provides positive aspects. Makes this makes no sense me. It doesn't give me value overall. I want the review. I want to fix potential bugs in the issues, yeah, and not like going for the positive things.''} (Phase I, P13).\\

\addlinespace

\multirow{6}{=}{Missing context} & \multirow{6}{=}{\emph{``Reading about stuff that is wrong''}} & \multirow{6}{=}{\emph{``[Less] practical''}} & \emph{``I think it's a little bit bloated, because it does like the context ... So I did feel like I was reading about stuff that is wrong, but it's not exactly wrong with the given context ... Some useful feedback there. But it's a little lost midst the bloat. It would be a little bit more practical if it was given by someone that already has the context''} (Phase I, P17).\\

\addlinespace

\multirow{6}{=}{Over-scoped suggestions} & \multirow{6}{=}{\emph{``too much of it [positives]''}} & \multirow{6}{=}{Beyond the scope of what \emph{``needs to be fixed''}} & \emph{``It does provide useful feedback. It's relevant exactly to the piece of code that I've written. It points out really well the positives and the negatives. But there's just too much of it. As a senior, I think I just want to know if there's anything that needs to be fixed''} (Phase I, P12).\\

\bottomrule

\end{tabular}
\end{table}

\textbf{Unaligned feedback}. Based on Phase I data, unaligned feedback is characterized by verbosity, generic tone, and lack of contextual precision. It introduces high cognitive demand rather than fluency. Table~\ref{tbl:rq4_unaligned} documents the elements of LLM feedback that caused disengagement. It seems that unaligned feedback disrupts cognitive flow. The misalignment with engineers' preferences leads to cognitive rejection. For example, in the case of P13 (Tbl.~\ref{tbl:rq4_unaligned}), his reaction seems to reflect a cognitive rejection triggered by goal, i.e., LLM's feedback misalignment. The LLM's feedback emphasized affirmation over actionable critique, so P13 experienced it as informationally redundant, leading to cognitive disengagement and reduced motivation to process or act on the comments.

While the Phase II prompt improved engagement and potential likelihood of adoption, a residual tension remained between trust in the LLM's accuracy and perceived superiority of human reviewers. Engineers valued the LLM for its completeness, precision, and emotional neutrality, yet several participants framed it as a supporting actor rather than a relational partner in review. As P19 reflected, the LLM's feedback is ``valuable'' but its ``faceless'' nature limits the shared understanding that sustains team-level accountability, \emph{``I think human feedback is still slightly more important, just for the fact that I know that I can give LLMs code that is super complex, and they will still understand while humans often do not ... I need to work with them, and I need my co-workers to understand my code too, they are still slightly more important for me when it comes to feedback, as compared to an LLM ... I still think LLMs pretty much always provide some valuable inputs, or at least even if it's missing context ... Well, ChatGPT being a faceless chat box is probably a factor''} (Phase II, P19). P15 similarly contrasted the LLM's technical completeness with the human warmth of peer investment: \emph{``so it found all the errors and all the things that I can improve. On the other hand, I can say I found the other participant review more, let's say earthworming, because another person spent time to read my code, understand it, and then do a review. I think I don't know like ChatGPT spend one second.''} (Phase II, P15). P7 captured this balance succinctly, describing the LLM not as a replacement but as \emph{``part of the toolbox''}, an aid that does not replace humans.

Collectively, these accounts suggest that while LLM feedback appeared to lower perceived cognitive effort, which engineers linked to a higher reported likelihood of adoption, it cannot replicate mutual accountability embedded in human peer review. In practice, some engineers envision a hybrid model of engagement, where LLMs augment human review by accelerating comprehension and surfacing issues, while humans preserve context, and shared responsibility for quality.

\bigskip

\begin{mdframed}[linewidth=0.5pt,backgroundcolor=gray!06]

\noindent\textbf{RQ4 -- Summary (LLM content \& Cognitive engagement)}\\

\noindent\textbf{Core finding.} Engineers reported higher engagement and a greater \emph{tendency} to adopt when LLM feedback \emph{aligns} with cognitive expectations (clear structure, concise scope, neutral tone, actionable guidance). Misalignment (verbosity, generic tone, missing context) increased processing effort and dampened engagement.\\

\noindent\textbf{Mechanism (reported in our data).} \emph{Feedback alignment} $\rightarrow$ \emph{cognitive ease} $\rightarrow$ \emph{improved engagement \& likelihood of adoption}. Alignment seems to reduce sense-making effort; however, misalignment created friction (high cognitive demand, mistrust).\\

\noindent\textbf{Boundaries.} Cognitive ease alone was not sufficient; engineers still prioritized \emph{perceived correctness} and \emph{context fit} before implementing.\\

\noindent\textbf{LLM-assisted vs. Peer-led review.} Emotional self-regulation demands seems to be lower; the locus of engagement skewed toward \emph{solo and internal sense-making} rather than dialogic negotiation.\\

\noindent\textbf{Residual tension.} Trust in LLMs and collective accountability remained stronger when engineers were asked to replace peers with LLMs. Several participants framed LLMs as \emph{tools} that assist but do not replace their peers. Most our participants see LLMs as first step to accelerate improvements and surface issues earlier; then, peers supply context, ensure norm enforcement, and shared accountability.\\

\noindent\textbf{Tooling design implications.} Our sample favors: (i) structured, sectioned outputs; (ii) concise, fit-for-purpose scope; (iii) neutral, depersonalized tone; (iv) brief why-based justifications; (v) context-aware suggestions that acknowledge uncertainty.\\

\end{mdframed}
\section{Discussion and Implications}
\label{sec:discussion}

\noindent We dedicate this section to discussing the theoretical and practical implications of our findings. Theoretically, our findings contribute by refining \emph{engagement theory}~\cite{kahn1990psychological,kahn2013relational,saks2014we,demerouti2001burnout,fredricks2004school}. We refine previous engagement models~\cite{kahn1990psychological} from framing emotional self-regulation as a preface to a co-evolving process with behavioral resolution, especially under dialogic regulation. This bi-directional relationship integrates affective regulation with behavioral resolution. We also extend existing theoretical frameworks by identifying social calibration as a relational control mechanism that conditions how engineers navigate the resolution process.

On the practical level, we identified a perceived shift in accountability in human-AI settings -- from individual and collective to individual only -- implying a reconfiguration of responsibility in AI-assisted software engineering (SE) work. Our findings further suggest that engineers experience AI-assisted reviews as diminishing certain socio-relational values 
(warmth, shared norms). Thus, we delineate the boundary conditions for effective human-AI collaboration in the context of SE.

\subsection{Theoretical Implications}

\noindent Kahn's theory explains the processes by which people adjust their ``selves-in-roles''~\cite{kahn1990psychological}. This theory's definition of engagement is well-aligned with our findings: \emph{``I defined personal engagement as the harnessing of organization members' selves to their work roles; in engagement, people employ and express themselves physically, cognitively, and emotionally during role performances''}~\cite{kahn1990psychological}. Kahn's theory provides a coherent analytical lens to interpret our high-level engagement model (\emph{emotional self-regulation $\rightarrow$ behavioral engagement $\rightarrow$ resolution and implementation}). His notion of the self being ``harnessed'' cognitively, emotionally, and physically aligns with the emotional readiness and subsequent behavioral moves we observed in the transition from and at the same time the intertwined relationship between \emph{feeling} and \emph{doing}. In addition, Kahn's theory is widely accepted and adopted in many fields of studies~\cite{may2004psychological,saks2006antecedents,rich2010job}, e.g., education~\cite{collie2012school}, healthcare~\cite{simpson2009engagement,hakanen2008job}, and service work~\cite{karatepe2013high,m2014linking}.

Kahn's theory premises that individuals can express both their personal selves and their roles effectively, allowing for a dynamic interaction where self and role inform each other. It emphasizes that personal engagement leads to authentic role behavior, showing one's thoughts, feelings, creativity, and personal connections while fulfilling job obligations~\cite{kahn1990psychological}. Our findings capture this relationship between ``self'' and role in the engineers' engagement trajectories. Through emotional self-regulation, engineers actively and unilaterally negotiate how much of their personal selves, i.e., emotions, values, and interpersonal climate, they are willing to bring into their professional role during code review. For example, those who choose reframing prefer internalizing emotion and prioritizing task resolution, whereas those who prefer dialogic regulation manage it interpersonally through interaction, allowing their authentic selves to surface in negotiation and meaning-making with their peers.

Behavioral engagement, in turn, represents the outward expression of that negotiation, where authenticity is balanced against role expectations, such as maintaining harmony, shared accountability, and team norms. In the case of LLM-assisted reviews, this interaction becomes more intrapersonal: engineers still enact authenticity through their commitment to quality and individual accountability for code quality, yet with fewer relational cues mediating the expression of self within the role. 

The tension we observed in the adoption of LLMs as reviewer may stem from this underlying role- identity conflict. When the social and relational components of the role are removed, engineers face a subtle threat to their professional identity~\cite{petriglieri2011under}: what it means to be a ``reviewer'' or to be ``reviewed'' becomes ambiguous. The LLM displaces the interpersonal dimension through which authenticity, recognition, and shared accountability are enacted. As a result, engagement with AI feedback is experienced less as a relational performance and more as an instrumental task. This shift may constrain the software engineer role and its expressive and identity-affirming aspects, challenging the equilibrium between the personal self and the professional role that Kahn's theory regards as central to authentic engagement~\cite{kahn1990psychological}.

Alami et al.~\cite{alami2025accountability,alami2024understanding} found that software engineers seek ``social validation'' through the quality of their code in code review. They pursue this ``validation'' from their peers by showcasing their personal standards in coding and building a reputation of a competent coder among their peers~\cite{alami2025accountability}. The introduction of the LLM in code review, in our study, may have also challenged this process and its underlying intents, conforming to group norms and reinforcing self-perception~\cite{alami2025accountability}.

\medskip

\begin{mdframed}[linewidth=0.5pt,backgroundcolor=gray!05,roundcorner=6pt,innertopmargin=6pt,innerbottommargin=6pt]

\noindent\textbf{Theoretical Implication 1: Reframing Engagement in Human-AI Contexts}\\

\noindent Our findings suggest that AI integration into socio-technical processes such as code review challenges the traditional alignment between the \emph{role} and the \emph{doing}. In peer-led reviews, engagement entails expressing the self through social interaction, negotiating meaning, sustaining norms, and receiving recognition, which reinforces authenticity in Kahn's sense of ``self-in-role''. When LLMs mediate or replace these interactions, the social grounding of work diminishes: engineers continue to \emph{do} the work (i.e., implement feedback), yet the \emph{role} through which they enact accountability, recognition, and belonging becomes ambiguous. 
This decoupling transforms engagement from a relational to an instrumental act, prompting engineers to question the authenticity of their contribution and the meaning of engagement itself in a context that is no longer human-to-human.

\end{mdframed}

\medskip

Kahn proposes three psychological conditions -- meaningfulness, safety, and availability -- that influence personal engagement in people's roles~\cite{kahn1990psychological}. The theory suggests that these conditions function as a contract between the person and their role, guiding their decision to engage or disengage.

\textbf{Psychological Meaningfulness.} Psychological meaningfulness is the ``sense of return on investments of self in role performances''~\cite{kahn1990psychological}. Kahn's theory claims that personal engagement is associated with psychological meaningfulness, and three factors influence this experience: task characteristics, role characteristics, and work interactions. Meaningful tasks are those that are challenging, varied, and allow for autonomy, yielding both competence and growth. Role characteristics involve identities that individuals must embody, which can align or conflict with their self-perception, affecting their sense of meaningfulness. Additionally, the status associated with roles contributes to feelings of value and recognition. Work interactions enhance psychological meaningfulness through rewarding relationships and mutual appreciation, which satisfy the need for relatedness. Conversely, negative interactions can diminish this sense of value, as respect and appreciation are key to preserving psychological meaning in one's work~\cite{kahn1990psychological}.

In our findings, meaningfulness emerges when engineers invest effort to demonstrate competence and pursue growth. Task characteristics are instantiated by the dual investment of emotional self-regulation and social calibration: reframing, dialogic clarification, and alignment to team norms to reduce friction and enable competent action, potentially leading to improvement (resolution and implementation). Role characteristics surface as behavioral engagement grounded in collective accountability. Engineers treat ``working toward resolution'' as part of the author role; enacting that role produces tangible returns (better code, learning), which are experienced as value and recognition (learning and codification). 

Work interactions complete the meaning loop: positive reinforcement, psychological safety, reciprocity of effort, and adaptive social comparison provide relatedness and legitimize the effort. Together, these mechanisms explain why peer-led review feels meaningful, even though socially and affectively effortful. Additionally, role obligations are affirmed through resolution, and relational exchanges validate contribution.

While our findings align with Kahn's work, we also bring nuances to the theory. Our findings show that engineers actively sustain meaning during code review, e.g., reframing criticism, dialoguing for clarity, aligning with team norms, and ensuring accountability. This behavior preserves professional integrity~\cite{alami2025accountability,khan2021understanding} and displays competence, previously recognized as psychological antecedents of meaningfulness in self-determination theory~\cite{ryan2000intrinsic}. Thus, in contrast to Kahn's original settings, where meaningfulness was embedded in inherently expressive (architecture) or service-oriented (camp workers) roles, our findings suggest that in evaluative and cognitively intense contexts such as code review, meaningfulness is actively sustained. Engineers achieve and preserve it through emotional and social investments to transform negative feedback into learning and improvements to the code. In this sense, meaningfulness in code review is both expression of self and preservation of self (worth and professional integrity) in the face of evaluation.

\medskip

\begin{mdframed}[linewidth=0.5pt,backgroundcolor=gray!05,roundcorner=6pt,innertopmargin=6pt,innerbottommargin=6pt]

\noindent\textbf{Theoretical Implication 2: Meaningfulness is both Expression and Preservation of Self}\\

\noindent In code review, meaningfulness arises when engineers invest emotionally and socially to transform evaluative exchanges into opportunities for learning and improvement of their code. Our findings extend Kahn's concept of psychological meaningfulness. We show that contextual, the task and the social obligations of the role shapes how engineers seek meaning. Therefore, it spans beyond the \emph{expression of self} to include the \emph{preservation of self}.

\end{mdframed}

\medskip

\textbf{Psychological safety.} Psychological safety (PS) is the third condition of engagement in Kahn's theory~\cite{kahn1990psychological}. It is the ability to express oneself without fear of negative repercussions~\cite{kahn1990psychological}. It is influenced by factors like supportive interpersonal relationships, group dynamics, management styles, and organizational norms. In environments that promote PS, individuals feel safer to engage. If they feel otherwise, they tend to withdraw~\cite{kahn1990psychological}. Our findings corroborate this condition. In peer-led code review, when engineers perceive the social climate as safe and non-punitive, they show willingness to reframe the feedback and/or initiate dialogue, reducing avoidance and defensive behavior. Therefore, PS seems to act as a social buffer that enables emotional self-regulation to transition into behavioral engagement. It may also shield against the negative effects during the resolution. Similar conclusions have been echoed in SE studies~\cite{alami2023antecedents,alami2024role,santana2025psychological,verwijs2023double,ahmad2025strengthening}.

\textbf{Psychological Availability.} ``Psychological Availability is the sense of having the physical, emotional, or psychological resources to personally engage at a particular moment''~\cite{kahn1990psychological}. The most relevant aspect of this condition to our findings is ``emotional energy'', which refers to the emotional labor required for individuals to invest in order to engage. Our findings further refine emotional labor as a process of \emph{self-regulation} that precedes behavioral engagement and may continue throughout the resolution process. Engineers expend emotional resources to down-regulate affect, reinterpret intent, and restore readiness to act constructively. This emotional labor is not arbitrary; it is value-driven and anchored in the commitment to code quality, accountability to peers, and a growth-oriented mindset. Our results thus extend Kahn's theory by showing how emotional labor can be simultaneously \emph{intrapersonal} (self-regulation of affect) and \emph{value-driven} (aligned with personal values).

\vspace{0.5cm}

\begin{mdframed}[linewidth=0.5pt,backgroundcolor=gray!05,roundcorner=6pt,innertopmargin=6pt,innerbottommargin=6pt]

\noindent\textbf{Theoretical Implication 3: Emotional Labor as Value-driven Psychological Availability}\\

\noindent Our findings extend Kahn's notion of psychological availability by framing emotional labor as a value-driven resource. Engineers mobilize emotional energy through self-regulation to transform evaluative tension into constructive action. This emotional labor is anchored in their commitment to code quality, accountability for their work and to their peers, and a growth mindset, which converts emotional expenditure into purposeful engagement. Psychological availability thus emerges as both intrapersonal and normative.

\end{mdframed}

\medskip

In software engineering (SE) research, emotions have attracted significant interest~\cite{novielli2019sentiment,sanchez2019taking}, e.g., in responding to requirements changes~\cite{madampe2022emotional}, in software artifacts~\cite{murgia2014developers}, detection~\cite{mahouachi2025enhancing}, and productivity of software developers~\cite{girardi2021emotions}. Theoretically, two key perspectives have emerged: one views emotions as continuous functions modeled across dimensions like valence (pleasantness/unpleasantness) and arousal (calm/excited)~\cite{russell1980circumplex}, while the other adopts a discrete approach identifying a limited set of basic emotions (e.g., joy, frustration)~\cite{ekman1999handbook}. Building on these perspectives, our evidence aligns with Gross's process model of emotion regulation~\cite{gross2015extended}.

The self-regulation repertoire we observed maps onto Gross's model of antecedent-focused strategies: \emph{situation selection} (anticipatory avoidance), \emph{situation modification} (dialogic clarification to surface intent and restore common ground), \emph{attentional deployment} (task-anchoring), and \emph{cognitive change} (reframing the intent); occasional \emph{response modulation} appears as defensiveness~\cite{gross2015extended}. In Gross's process model, emotions are assumed to arise from ongoing appraisals of situations~\cite{gross2015extended}. In our findings, appraisal is value-driven: engineers evaluate comments against goals (quality), norms (accountability, peer harmony), and mindset (growth), so self-regulation is not merely hedonic~\cite{gross1998emerging,gross2015extended} (down-regulating unpleasant affect) but eudaimonic~\cite{gross1998emerging,gross2015extended} (pleasure achieved through a meaningful life), aimed at preserving competence and relatedness.

While the introduction of LLM-assisted review appears to reduce some self-regulation demands, it stripped interpersonal appraisal cues (lower arousal, reduced relatedness) and shifted resolution toward individual sense-making. Yet the LLM cannot supply the relatedness and shared accountability that sustain team-level meaning.

\subsection{Practical Implications \& Future Research}

Our findings bring implications to the individual level (i.e., software engineers) and human-AI collaboration in the context of software engineering. At the individual level, we contend that we cannot fully engineer out harshness and negativity from such a process; they recur in interpersonal evaluation processes. The pragmatic lever is to equip engineers with self-awareness and agency skills so they can preserve and achieve meaningfulness despite imperfect peers' behavior. Our findings also demonstrate that in a socio-technical process such as code review, the complex trajectory to resolution is intertwined with seeking meaningfulness. Therefore, in similar processes, AI should act as a supportive accelerator rather than a social substitute. At the organizational level, the intricate relationship between software engineers' role identity and their tasks uncovers a deeply social, value-laden sense of meaning and accountability that the adoption of AI may challenge.

\subsubsection{Individual Level}

Toxicity has been studied in code review~\cite{sarker2023toxispanse,egelman2020predicting} and open source~\cite{carillo2016towards,miller2022did,sarker2025landscape} extensively in SE research~\cite{imran2025silent}. While the work improved our understanding of its causes, detection, and impacts, collectively the literature does not frame it as something to eliminate, but rather a reality that should be managed and mitigated~\cite{seering2017shaping}. For example, Alami et al.~\cite{alami2019does} found that despite the negativity experienced in open-source code reviews, contributors' intrinsic and extrinsic motivation enabled them to sustain their contributions, suggesting that interventions should cultivate value-based agency and coping skills.

Practically, we argue that our findings imply that interventions should target software engineers' self-regulation skills and value clarity rather than attempting to eliminate negative affect from peer review. We recommend targeting \emph{education}~\cite{panadero2017review,durlak2011impact}, \emph{onboarding}~\cite{saks2007socialization,bauer2007newcomer}, and \emph{mentoring}~\cite{eby2013interdisciplinary} because they align with offering effective avenues for learning~\cite{panadero2017review,bauer2007newcomer,eby2013interdisciplinary}. For instance, education provides scalable, low-stakes, and early exposure opportunities for teaching self-regulation and value clarity~\cite{durlak2011impact}. Onboarding is a high-leverage socialization window~\cite{bauer2007newcomer}, where norms, accountability expectations, and review etiquette are most influenced at entry and potentially can shape behaviors. Mentoring provides the situated, relational reinforcement that a classroom or policy cannot~\cite{eby2013interdisciplinary}.

\begin{itemize}

\item \textbf{Education.} \emph{How:} embed short modules on self-regulation, value clarity (quality, accountability, peer harmony), and dialogic skills within SE curriculum and professional development.

\item \textbf{Onboarding.} \emph{How:} add a review readiness track to developer onboarding: (i) norms for psychological safety and reciprocity of effort; (ii) tutorial on personal engagement plan (what triggers emotions, preferred self-regulations, escalation path).

\item \textbf{Mentoring.} \emph{How:} pair newcomers with a review mentor; use brief, scheduled debriefs after challenging reviews to surface regulation choices (``what I felt?, what I valued?, what I did?'') and to align with team norms. Encourage mentors to model value-based reasoning (why a change matters for quality, accountability, and meaningfulness).

\item \textbf{Team practices.} \emph{How:} institutionalize short, recurring touchpoints (e.g., monthly 15-minute review health check) to re-affirm norms, share effective self-regulation moves, and calibrate expectations.

\end{itemize}

\noindent \textbf{\emph{Future research.}} Future research may consider evaluating which delivery modes (course module vs.\ onboarding vs.\ mentoring) are most effective at improving perceived meaningfulness and reducing avoidance and defensive responding.

\subsubsection{Human-AI Collaboration in Software Engineering}

Practically, our findings suggest that LLM integration should be positioned as a \emph{supportive co-reviewer} that preserves the social sources of meaningfulness (team norms, shared accountability, negotiated understanding) rather than replacing them. We also reported that when engineers deal with LLM alone, their decisions become unilateral. While the engagement with the AI tool lowers emotional cost, it also emphasizes the move from emotion self-regulation to \emph{ethics, value alignment, accountability, and governance literacy}. Interventions should therefore target the same delivery channels: \emph{education}~\cite{panadero2017review,durlak2011impact}, \emph{onboarding}~\cite{saks2007socialization,bauer2007newcomer}, and \emph{mentoring}~\cite{eby2013interdisciplinary}.

\begin{itemize}

\item \textbf{Education.} \emph{How:} embed short, practice-based modules on (i) AI affordances and limits (hallucinations, context gaps, evaluation criteria), (ii) value alignment and professional accountability when acting on AI suggestions, and (iii) introduce AI governance practices.

\item \textbf{Onboarding.} \emph{How:} add an AI-in-your-work track: (i) organizational rules for when and where AI may be used; (ii) verification and context-enrichment steps before adoption; (iii) accountability alignment (who is responsible when AI contributes).

\item \textbf{Mentoring.} \emph{How:} mentors should model judgment with AI use: reading LLM feedback/AI outputs critically, articulating why suggestions are accepted or rejected, when to escalate to team-level verification and approval.

\item \textbf{Team practices.} \emph{How:} institutionalize hybrid review rituals that keep humans in the loop: (i) define when peer sign-off is required for AI-influenced changes, e.g., security/architecture; (ii) schedule a monthly AI review health check to align AI-assisted practices with norm compliance, correctness, and downstream rework. 

\item \textbf{Tooling \& policy.} \emph{How:} Publish and socialize an AI usage policy scoped to software engineering workflows (permitted use, prohibited use, data handling, escalation).

\end{itemize}

\noindent \textbf{\emph{Future research.}} Future research may consider evaluating (a) whether AI hybrid adoption (human sign-off $+$ AI triage) preserves perceived meaningfulness and collective accountability compared to AI-only triage in code review; (b) trade-offs between AI adoption and team-level cohesion; and (c) how governance literacy (ethics, accountability) moderates trust when AI participation scales.

\subsubsection{AI Adoption in Software Engineering Practices}

Our findings suggest that when AI is introduced into socially meaningful practices such as code review, it not only adds a new tool, as historically has been the case, e.g., static analyzers, CI services, etc.; but it also reconfigures and challenges what it means to be a software engineer. Based on Kahn's theory~\cite{kahn1990psychological}, in the case of peer-led review, engineers enact roles that are explicitly social. They give and receive recognition, negotiate standards, and share accountability. However, the introduction of an LLM has narrowed this social and human context, from which engineers derive a sense of professional identity and meaningfulness from interaction with peers, validation through the quality of their code, and reciprocal evaluation.

AI adoption strategies in SE should therefore be preferably framed as \emph{role-supporting} rather than \emph{role-replacing}. Concretely, organizations should define AI as a tool that scaffold and extend human expertise. Making the AI software engineer role boundaries explicit in policies, onboarding, and everyday discourse may reduce the threat to role identity and maintain the sense that AI is there to support engineers in doing \emph{their} work, rather than hollowing it out. 

Alternatively, as a research community, we need to further investigate the role identity of software engineers, unpack the role identity and individual relationship, and how both the human and social contexts nurture this identity, as well as how AI can become a partner in an environment where meaning is derived from social interactions.

At the same time, when decisions on LLM-generated feedback become more unilateral and less socially mediated, the future skill of software engineers should emphasize ethical discernment, value alignment, and accountability when acting on AI advice. Education, onboarding, and mentoring should therefore integrate not only technical skills to use AI but also governance literacy; when to trust or challenge AI outputs, how to document AI influence on changes, and how to escalate AI-shaped decisions to team-level discussion when values, norms, or safety are at stake.

Our findings also show potential promises. Our experiment showed that aligning the feedback format with the engineers' preferences reduces cognitive demands, thereby increasing the likelihood of adoption. We also report lower emotional and social costs when developers processed LLM-assisted feedback.  Nonetheless, our findings explain why LLMs should be adopted thoughtfully and carefully in the code review process. Developers’ emotional engagement with human-authored comments is a crucial step in their sense-making process, and the social nature of code review, combined with the lack of engagement with LLM-generated feedback, means that naive adoption risks opening Pandora’s box, changing how developers process and resolve feedback, and sidestepping a detailed and intricate sense-making process. To preserve the depth of the code review process, we recommend the following two adoption strategies: \textbf{AI-primed review} (LLM as first pass) and \textbf{AI-mediated review} (LLM as an emotional buffer for peer feedback).

For AI-primed review, the AI tool performs a first-pass review on the code. The aim of this early and preliminary review is to surface issues and suggestions before a second review by peers. This type of evaluation may reduce errors and prepare the code for a second peer-led review, where the focus may shift to contextual issues and higher-order concerns such as design trade-offs, team standards, and shared understanding of the code rather than surface-level errors. For the AI-mediated review, AI could be integrated into the review workflow to rephrase, structure, and improve the tone of peers' feedback to reduce the emotional burden on the authors.

\medskip

\noindent \textbf{\emph{Future research.}} Future research may consider investigating: (a) how AI-assisted software engineering practices affect software engineers' role identity and perceived meaningfulness of work; (b) what governance mechanisms (policies, workflow constraints, and oversight practices) may effectively ensure accountability, transparency, standards, and norm compliance when AI tools participate in software engineering decision-making; and (c) how software engineering education can integrate AI-assisted tools in ways that foster AI collaboration literacy while preserving students' foundational SE skills (e.g., programming, design, debugging, and code comprehension).
\section{Research Trustworthiness}
\label{sec:trust}

We implemented several techniques to address the requirements of research trustworthiness~\cite{miles2014qualitative}. In this section, we report the techniques we used: \textit{Saturation}, \textit{Member checking}, \textit{Peer debriefing}, and \textit{Thick description}.

\textit{Saturation}: We triangulated data sources, including interviews, focus groups, and participant feedback sessions. This exercise allowed us to ensure that our findings are corroborated across different data sources and contexts.

As discussed in Sect.~\ref{sec:methods}, during the re-analysis of the data Phase I then II, we monitored thematic \emph{saturation}~\cite{morse2004theoretical,aldiabat2018data,miles1984qualitative,hennink2022sample}. We documented the outcome of the monitoring process and made the spreadsheet available in our shared documents package (see Sect.~\ref{sec:replication}).

\textit{Member checking}: Although we conducted a member checking activity in our previous work~\cite{alami2025human}, upon the completion of the analysis, we organized a second \emph{member checking}~\cite{birt2016member} activity to collect feedback from our interviewees on our findings. We summarized our RQs' findings in an online questionnaire and invited all 20 participants to participate. First, we asked them for each finding to either ``agree,'' ``neither agree nor disagree,'' or ``disagree." Then, irrespective of their level of agreement, we asked them to comment on the findings. Table~\ref{tbl:member_checking} summarizes the outcome of our member checking and documents some illustrative comments. Sixteen participants responded. The data and the questionnaire we used are available in our shared documents package (see Sect.~\ref{sec:replication}).

\textit{Peer debriefing}: Although the analysis was primarily conducted by the first author, the second and fifth authors reviewed the proposed codes, and the results were continuously discussed and scrutinized by the other two authors in several meetings throughout the analysis process. The participation of two authors in the coding process helped minimize researcher biases~\citep{miles2014qualitative}. This approach is grounded in our epistemological stance, constructivism, which posits that knowledge is socially constructed and that collective intellectual engagement can lead to more reliable understandings of the data~\cite{miles2014qualitative}.

\textit{Thick description}: We endeavored to provide a detailed explanation of our research process and the decisions we have made throughout (see Sect.~\ref{sec:methods}). In addition, we assembled a comprehensive replication package (see Sect.~\ref{sec:replication}).

\begin{table*}[t!]
\footnotesize

\caption{Member checking summary by research question (RQ2--RQ4). We invited all 20 participants; only participants who took part of Phase II could answer and comment of RQ4 (two participants took part of the member checking but not the followup interviews).}
\label{tbl:member_checking}

\renewcommand\arraystretch{1.0}
\setlength{\tabcolsep}{5pt}

\begin{tabularx}{\textwidth}{l ccc X}
\toprule

\multirow{2}{*}{\textbf{RQ}} &
\multicolumn{3}{c}{\textbf{Level of agreement (count)}} &
\multirow{2}{*}{\textbf{Illustrative participant comments}} \\
\cmidrule(lr){2-4}
& \textbf{Agree} & \textbf{Neither} & \textbf{Disagree} & \\

\midrule

\textbf{RQ2} &  15  &  1  &  --  &

\emph{``When it comes to myself and human feedback, I usually look towards dialogic regulation. Feedback, when written down, can come over more harsh than people speaking to each other. I have often found myself in a situation where the feedback sounded harsh, but talking about it showed me that there was actually no bad blood involved''} (P19).\\

\addlinespace

\textbf{RQ3} &  16  &  --  &  --  &

\emph{``The interpretation makes sense, and the steps list above makes sense. Developers usually tend to conform after they received feedback when its negative. They will also start paying attention code related team culture, team practices if one exists otherwise raise an attention to get those standards and practices incorporated into the team. The bit on LLM also makes sense, they can really speed up tasks but usually lacks context and unable to anticipate certain scenarios. Although with the advancement with todays LLM's these seem to be changing''} (P3).\\

\addlinespace

\textbf{RQ4} &  14  &  --  &  --  &

\emph{``Yes, I agree that the overly verbose LLM feedback was not needed, from the LLM I only need what is wrong with the code and how to fix it. I would be more receptive to more detailed feedback and reasoning from a human''} (P13).\\

\bottomrule

\end{tabularx}
\end{table*}

\section{Limitations and Trade-offs}
\label{sec:limit}

We conducted the code review process anonymously. We were constrained by the Prolific requirement to maintain participant anonymity throughout the study, which is not the case in open source and proprietary software development. In the latter, developers are colleagues and interact directly on a daily basis. In such a social context, developers may align the tone of their feedback according to status, relationships, and team culture~\cite{bird2006mining,kononenko2016code}. Similarly, in open-source development, communities adopt different styles and strategies in pull request reviews, from ``lenient'' to ``protective'' according to their history, culture, and sustainability strategy~\cite{alami2020foss,alami2021pull}. Aware of this limitation, we asked during the interview whether this factor has influenced our participants approach to reviews. Most participants claimed this was not the case.

The code reviews were conducted in an artificial setting characteristic of a research environment. Such a controlled environment may not capture the inherent complexity of a real-world setting. However, this controlled environment allowed us to focus on the variables of interest and gain nuanced findings~\cite{stol2018abc}. We also acknowledge that our participants may have modified their approach to the review compared to a professional setting, given the research nature of their contribution. However, in the interview, we prompted all our participants to explain the content and the delivery style they used. This ``reflective questioning'' uncovered the ``authentic'' reasoning even when the original actions occurred in artificial settings~\cite{silverman2013counts}.

We opted for ChatGPT 4o as the LLM of choice for the study due to its wide accessibility and familiarity among a diverse group of participants. This model may not represent the full spectrum of available LLMs. Different models may generate varying quality of feedback, which may impact participants' perceptions and engagement with the reviews. However, ChatGPT is one of the state-of-the-art LLM tools, which makes it a timely and relevant choice. In addition, using a single LLM allowed us to collect consistent data, serving as a baseline for future research.

Another methodological limitation characteristic of interview studies is response bias~\cite{podsakoff2003common}. We remediated this potential limitation by anchoring the interview guide in the reviews received by the interviewees from other participants. This strategy enhanced the relevance and contextual accuracy of the data we collected.

We also did not explore other factors outside the team's context. Our study setup remained constrained with the context of a team; organizational factors such as career advancement and financial rewards could also influence engineers' behaviors in how they deal with feedback. For example, while the self-regulation strategies seem to be motivated by intrinsic drives such as personal motivation for quality and a growth mindset, extrinsic drives may also influence engineers' inclination to adopt strategies better aligned with organizational expectations. Alami \& Ernst~\cite{alami2024understanding} found that software engineers' accountability for code quality is also influenced by institutional drives, such as financial rewards and performance evaluation.

\section{Conclusion}
\label{sec:conclusion}

In this study, we framed code review as a socio-technical practice, and we sought to understand engagement in the context of peer-led and LLM-assisted reviews. We found that engineers move from feeling to doing: emotional self-regulation $\rightarrow$ behavioral engagement (social calibration and resolution) $\rightarrow$ outcomes (adoption, learning, and norm codification). By comparing this loop in peer- and LLM-assisted settings, we found that emotional self-regulation and social calibration become less relevant. We also identified an accountability shift when AI enters the workflow: from shared, peer-enforced responsibility toward individuals. We extend engagement theory~\cite{kahn1990psychological} beyond ``expression of self in role'' to include the \emph{preservation of self} under evaluation conditions (e.g., code review), and we surface social calibration as a relational control mechanism that conditions how resolution unfolds. We also inform the theory by demonstrating that emotional labor is value-driven.

Practically, while we cannot engineer negativity out of human review, we can equip engineers to manage it. Meaningfulness is sustained not by sterilizing emotion, but by strengthening self-regulatory repertoires and clarifying values, accountability, and peers' expectations through education, onboarding, mentoring, and team practices. We also recommend integrating LLMs as supportive co-reviewers without displacing the social sources of meaning, to minimize disruption to the social integrity of code review. Hybrid practices and human sign-off on AI-influenced changes are potential also adoption avenues.

In sum, we report that the path to a higher-quality code runs through human judgment made sturdier by values and tools that know their places. We recommend AI for support, keeping the human in the loop for truth, accountability, and meaning.

%% The acknowledgments section is defined using the "acks" environment
%% (and NOT an unnumbered section). This ensures the proper
%% identification of the section in the article metadata, and the
%% consistent spelling of the heading.

\begin{acks}

We would like to thank our participants for their time and effort in making this study possible. This work is supported by the Innovation Fund Denmark for the project AI4SE1DK (4354-00006B).

\end{acks}

%%
%% The next two lines define the bibliography style to be used, and
%% the bibliography file.
\bibliographystyle{ACM-Reference-Format}
\bibliography{references}

\end{document}